\documentclass[10pt]{article}

\usepackage{amsmath}
\usepackage{amssymb}
\usepackage[margin=0.8in]{geometry}
\usepackage{xcolor}
\usepackage{graphicx}
\usepackage{hyphenat}
\usepackage{bm}
\usepackage{siunitx}
\usepackage{enumerate}
\usepackage{abstract}

\usepackage[superscript]{cite}

\usepackage{draftwatermark}
\SetWatermarkText{\bf PREPRINT ~~~ PREPRINT ~~~ PREPRINT   ~~~ PREPRINT}
\SetWatermarkFontSize{0.8cm}
\SetWatermarkAngle{90}
\SetWatermarkHorCenter{20.5cm}
\SetWatermarkColor[gray]{0.75}

\title{Numerical simulation of rarefied supersonic flows using a fourth-order maximum-entropy moment method with interpolative closure}

\author{Stefano Boccelli\thanks{Corresponding author: stefano.boccelli@polimi.it}\\
\textit{\normalsize University of Ottawa, ON, Canada,}\\
\textit{\normalsize and Politecnico di Milano, Milan, Italy,}\\
\textit{\normalsize and von Karman Institute for Fluid Dynamics, Sint-Genesius-Rode, Belgium.}\\[2ex]
Willem Kaufmann\\
\textit{\normalsize University of Ottawa, ON, Canada.}\\[2ex]
Thierry E.~Magin\\
\textit{\normalsize von Karman Institute for Fluid Dynamics, Sint-Genesius-Rode, Belgium.}\\[2ex]
James G.~McDonald\\
\textit{\normalsize University of Ottawa, ON, Canada.}
}

\date{\textbf{Published version: Journal of Computational Physics, 2024}\\ DOI: 10.1016/j.jcp.2023.112631}

\begin{document}

\maketitle

\begin{abstract}
Maximum-entropy moment methods allow for the modelling of gases from the continuum regime to strongly rarefied conditions.
The development of approximated solutions to the entropy maximization problem has made these methods computationally affordable.
In this work, we apply a fourth-order maximum-entropy moment method to the study of supersonic rarefied flows.
For such conditions, we compare the maximum-entropy solutions to results obtained from the kinetic theory of gases at different Knudsen numbers.
The analysis is performed for both a simplified model of a gas with a single translational degree of freedom (5-moment system)
and for a typical gas with three degrees of freedom (14-moment system).
The maximum-entropy method is applied to the study of the Sod shock-tube problem at various rarefaction levels, and to the simulation of 
two-dimensional low-collisional crossed supersonic jets.
We show that, in rarefied supersonic conditions, it is important to employ accurate estimates of the wave speeds.
Since analytical expressions are not presently available, we propose an approximation, valid for the 14-moment system. 
In these conditions, the solution of the maximum-entropy system is shown to realize large degrees of non-equilibrium and to approach the Junk subspace, 
yet provides a good overall accuracy and agreement with the kinetic theory.
Numerical procedures for reaching second-order accurate discretizations are discussed, as well as the implementation of the 14-moment solver on Graphics Processing Units (GPUs).
\end{abstract}

\section{Introduction}\label{sec:sample1}

The development of mathematical and numerical models for simulating rarefied supersonic flows is of relevance to a variety of aerospace problems.
These include the study of the atmospheric entry of spacecrafts and meteors \cite{schouler2021ixv,josyula2015hypersonic,bariselli2018aerothermodynamic}, 
the design of air-breathing electric propulsion systems \cite{romano2022design},
and low-density aerodynamic facilities \cite{oiko2019roar}.
The characteristic feature of such flows is a relatively slow collisional relaxation time, often slower or comparable to
the advection time scales. 
As a result, the distribution that characterizes the particle velocities can be non-Maxwellian, and one obtains
non-equilibrium effects such as temperature anisotropy and non-local transport \cite{ferziger1972mathematical}.
In such situations, the classical fluid and hydrodynamic formulations 
(the Euler and Navier-Stokes-Fourier (NSF) equations of gas dynamics)
may predict incorrect transport processes, resulting in inaccurate or even unphysical results \cite{josyula2015hypersonic,lofthouse2007effects}.
Analogous problems appear in the presence of strong spacial gradients, comparable to the collisional mean free path,
a typical case being the study of the internal structure of shock waves \cite{pham1989nonequilibrium}.

Accurate simulations of non-equilibrium flows can be obtained through the kinetic theory of gases \cite{cercignani1988boltzmann},
solving the Boltzmann equation for the velocity distribution function (VDF), either with a direct numerical discretization of the phase space \cite{mieussens2000discrete} or through the use of
particle methods \cite{bird1994molecular}.
Besides the computational cost associated with the higher dimensionality of the kinetic equation, 
these methods become increasingly computationally demanding as 
one increases the collisionality and enters the transitional regime.
This problem has triggered the development of moment methods, as a 
computationally cheaper alternative to the full Boltzmann equation, 
with, at the same time, a superior accuracy with respect to the 
traditional Euler/NSF formulations \cite{torrilhon2016modeling}.
The approach consists in deriving a set of governing equations by computing statistical moments of the kinetic equation \cite{struchtrup2005macroscopic}.
Moments are weighted averages of the particle distribution function and correspond to macroscopic quantities such as the density, momentum, and energy of the gas. 
As higher-order moments are considered, one can hope to obtain a more accurate formulation.
Among the most common moment methods, we should cite the Grad method \cite{grad1949kinetic} and its regularizations \cite{struchtrup2003regularization,cai2014globally}, 
quadrature methods \cite{fox2009higher,van2021higher} and maximum-entropy methods \cite{levermore1996moment,muller1993extended}.
These methods differ by the way in which the moments of the highest order, that appear in the convective fluxes, 
are approximated (or ``closed'').

The maximum-entropy family of moment methods possesses a number of desirable properties, such as global hyperbolicity, and guarantees a positive VDF by construction. 
These methods happen to be robust, even in strong non-equilibrium conditions.
However, the widespread employment of such methods has been hindered by their high computational cost, 
associated with the need to maximize the entropy numerically, in every cell of the computational domain.
This issue has been relieved by the development of approximated interpolative solutions to the entropy maximization problem \cite{mcdonald2013affordable,mcdonald2016approximate,giroux2021approximation}, which has made a selection of these methods computationally affordable.
In the past, the maximum-entropy formulation, either with the standard or the approximated interpolative closure, 
has been applied to a number of problems, including the study of instabilities and turbulence \cite{kaufmann2022largescale,ivan2022directnumerical},
radiation transport \cite{sarr2020second} and the electron and ion dynamics in plasmas \cite{boccelli202014,boccelli202214}.

The aim of this work is to investigate the application of fourth-order maximum-entropy methods to rarefied conditions, up to free-molecular flows, 
and in the presence of supersonic flow features.
First, in Section \ref{sec:max-ent}, we review the formulation of these methods for one-dimensional and for three-dimensional gases, starting from kinetic theory.
The numerical solution of the maximum-entropy systems is discussed in Section \ref{sec:numerical-solution}, 
considering a time-explicit finite-volume formulation, both on a single processor and on Graphics Processing Units (GPUs).
In Section \ref{sec:sod-shock}, we study a one-dimensional gas: the maximum-entropy method is applied to the study of the Sod shock-tube problem, 
at different levels of collisionality, from the continuum to the free-molecular regime.
The resulting wave structure and trajectory in moment space are discussed.
The same test case is then employed in Section~\ref{sec:approx-wave-speeds} to analyze the wave speeds of the system.
In Section~\ref{sec:approx-wave-speeds}, we also propose an empirical approximation for the maximum and minimum wave speeds of the 14-moment system.
This approximation allows for a considerable reduction in the computational cost of the simulations.
A full derivation is discussed in \ref{appendix:approx-wave-speeds}.
Then, in Section \ref{sec:two-jets}, we consider a two-dimensional problem consisting of either a single collisionless jet expanding into a vacuum, or 
two rarefied jets crossing each other, at various collisionalities. 
The solution is compared to kinetic solutions.
A solution of the Euler equations for this problem is discussed in \ref{sec:appendix-euler-jets}.


\section{The fourth-order maximum-entropy method}\label{sec:max-ent}

In this section, we sketch the derivation of the maximum-entropy system of
equations, starting from kinetic theory.
For a full derivation and a detailed discussion of the method, 
the reader is referred to \cite{levermore1996moment}.

\subsection{The kinetic equation}

In gas-kinetic theory, the state of a single-species gas is described by the particle velocity distribution function (VDF), 
$f(\bm{x},\bm{v},t)$, that depends on the position, $\bm{x}$, on the particle velocity, $\bm{v}$, and time, $t$.
For a single-species neutral gas, in absence of external forces, the evolution equation reads \cite{ferziger1972mathematical}
\begin{equation}\label{eq:boltzmann-bgk-eq}
  \frac{\partial f}{\partial t} + \bm{v} \cdot \frac{\partial f}{\partial \bm{x}} = \mathcal{C}(f) \, ,
\end{equation}

\noindent where $\mathcal{C}(f)$ is the collision operator.
For neutral gases, one should employ the Boltzmann collision operator.
However, the application of this operator to the maximum-entropy moment equations is not trivial.
Therefore, we opt here for a simpler model, 
the BGK collision operator \cite{bhatnagar1954model},
that can be easily implemented in moment systems, and thus allows us to obtain 
a consistent comparison between kinetic and fluid models.
We write
\begin{equation}
  \mathcal{C}(f) = - (f - \mathcal{M})/\tau \, ,
\end{equation}

\noindent where $\mathcal{M}=\mathcal{M}(\bm{x},\bm{v},t)$ is a Maxwellian VDF evaluated at the local density, bulk velocity, and temperature,
while $\tau$ is the mean collision time, equal to the inverse of the velocity-independent collision frequency.
Among the drawbacks of the BGK collision operator, it should be mentioned that it does not predict the correct Prandtl number.
Since the collision frequency is constant, all moments of the distribution function relax at the same rate.
These assumptions are not justified in supersonic rarefied flows.
Yet, its broad usage makes the BGK model an established benchmark.
Also, in this work, we mainly aim at analyzing how well maximum-entropy methods can approximate the streaming operator 
(the left-hand side of the kinetic equation).
Consequently, we regard an accurate modelling of collisions as secondary.

Macroscopic quantities, describing the gas state, are obtained as statistical moments of the VDF.
The moment, $M_\psi(\bm{x}, t)$, associated with the particle quantity $\psi$ is
\begin{equation}
  M_{\psi}(\bm{x}, t) = \left< \psi f \right> = \iiint_{-\infty}^{+\infty} \psi f(\bm{x}, \bm{v}, t) \, \mathrm{d}^3 v \, .
\end{equation}

For simplicity, in the following, we drop the spacial and temporal dependencies.
The gas density, average momentum, and the pressure tensor components are obtained respectively as
\begin{equation}
  \rho = \left< m  f \right> \ \ , \ \ \ \rho u_i = \left< m v_i \, f \right> \ \ , \ \ \ P_{ij} = \left< m c_i c_j\, f \right> \, ,
\end{equation}

\noindent where $m$ is the particle mass and $c_i = v_i - u_i$ is the peculiar velocity along $i=\{x,y,z\}$.
The hydrostatic pressure is defined as $P = (P_{xx}+P_{yy}+P_{zz})/3$.
One can define moments up to an arbitrary order in the velocity.
In this work, we also need the following moments:
the components of the heat-flux tensor, $Q_{ijk}$, the fourth-order moment tensor, $R_{ijkl}$, and the 
fifth-order tensor, $S_{ijklm}$, defined as 
\begin{equation}
  Q_{ijk} = \left< m c_i c_j c_k \, f \right> \ \ , \ \ \ R_{ijkl} = \left< m c_i c_j c_k  c_l \, f \right> \ \ , \ \ \ S_{ijklm} = \left< m c_i c_j c_k  c_l c_m \, f \right> \, .
\end{equation}

Moments obtained about the peculiar velocity are denoted as central moments. 
Such moments are free from the effect of advection and only depend on 
(i) the density and (ii) the detailed shape of the distribution function.
In the following, repeated indices imply summation.
The heat-flux vector is the contraction of the heat-flux tensor, $q_i = Q_{ikk}$.
Note that this definition differs from the typical fluid dynamic definition by a factor of $1/2$.
Being proportional to an odd power of the velocity, the heat flux is associated with the asymmetry of the
VDF.

The fourth-order moment, $R_{ijkl}$, involves the product of four peculiar velocity components.
Its contraction, $R_{iijj}$, is proportional to the particle velocity to the fourth power, and is associated with the 
kurtosis of the VDF.
To fix the ideas, $R_{iijj}$ assumes a super-Maxwellian value if particles are displaced from the bulk of the VDF towards the tails, and is sub-Maxwellian in the opposite case.
In the moment method considered in this work, the fifth-order moment tensor, $S_{ijklm}$, is a closing term.
Its treatment is discussed in Section~\ref{sec:phys-real-junk-closing-1D1V}.

For a Maxwellian VDF, $Q_{ijk}$ and $R_{iijj}$ assume the equilibrium values:
$Q_{ijk}^{\mathrm{eq}} = 0$ and $R_{iijj}^{\mathrm{eq}} = 15 P^2/\rho$.
The moments of the VDF can be non-dimensionalized through a scaling by suitable powers of the density and the characteristic speed, $\sqrt{P/\rho}$,
\begin{equation}\label{eq:dimensionless-variables}
  \begin{cases}
    \rho^\star = 1 \, , \\
    P_{ij}^\star = P_{ij}/P \, , \\
    Q_{ijk}^\star = Q_{ijk}\Big/\left( \rho \sqrt{P/\rho}^{\, 3} \right) \, , \\
    R_{ijkl}^\star = R_{ijkl}\Big/\left( \rho \sqrt{P/\rho}^{\, 4} \right) \, . \\
    S_{ijklm}^\star = S_{ijklm}\Big/\left( \rho \sqrt{P/\rho}^{\, 5} \right) \, . \\
  \end{cases}
\end{equation}

One can obtain evolution equations for the moments by taking averages of the kinetic equation.
Following \cite{ferziger1972mathematical} and using the BGK collision operator, 
\begin{equation}\label{eq:ch-theory-general-equation-moments}
  \frac{\partial \left< \psi f \right>}{\partial t}
+ \frac{\partial}{\partial \bm{x}} \cdot \left< \bm{v}\, \psi  \, f \right>
= 
- \frac{\left< \psi f \right> - \left< \psi \mathcal{M} \right>}{\tau} \, .
\end{equation}

\noindent After identifying a set of moments of interest, $\bm{U}$, generated
by a set of particle quantities, $\psi$, and after applying
Eq.~\eqref{eq:ch-theory-general-equation-moments} to each of them, one obtains a system of partial differential equations in balance-law form,
\begin{equation}\label{eq:PDEs-balance}
  \frac{\partial \bm{U}}{\partial t}
+ \bm{\nabla} \cdot \bm{F}
= 
  \bm{C} \, ,
\end{equation}

\noindent where $\bm{F} = [\bm{F}_x, \bm{F}_y, \bm{F}_z]$ are spacial 
fluxes, and $\bm{C}$ contains the collisional source terms.
This system is not closed, since some entries in the fluxes need to be further specified.
Indeed, while Eq.~\eqref{eq:ch-theory-general-equation-moments} describes the evolution of a given moment, $\left<\psi f\right>$,
space fluxes include unspecified terms of higher order in the velocity, $\left< \bm{v} \,\psi f\right>$. 
The closure of this system is discussed in the next section.

\subsection{The maximum-entropy closure}

The maximum-entropy assumption \cite{levermore1996moment} provides a way to close the system shown 
in Eq.~\eqref{eq:PDEs-balance}.
First, it should be recalled that, while solving Eq.~\eqref{eq:PDEs-balance}, one only knows the value of the state vector, $\bm{U}$, and does not know the shape of the associated distribution function.
Indeed, since $\bm{U}$ is composed of a finite number of moments, there could be an infinite number of distribution functions that are compatible with
such moments.
The maximum-entropy assumption prescribes that: \textit{among all possible distribution functions, $f$, that integrate to $\bm{U}$, one should select the one associated with the highest probability.}
For a classical ideal gas, the most probable distribution function can be shown to be the one that maximizes the statistical entropy \cite{muller1985thermodynamics}, defined as
\begin{equation}
  \mathcal{S} = - k_B \iiint f \ln \frac{f}{y} \, \mathrm{d}^3 v \, ,
\end{equation}

\noindent where $y$ is a scaling parameter and $k_B$ is the Boltzmann constant.
For a classical gas, the maximum-entropy distribution function can be shown to take the shape
\begin{equation}\label{eq:max-ent-VDF-definition-exp}
  f_{\mathrm{M.E.}} = \exp(\bm{\alpha}^\intercal \bm{m}(\bm{v})) \, ,
\end{equation}

\noindent where $\bm{\alpha} = \bm{\alpha}(\bm{x}, t)$ is a vector of parameters, while $\bm{m}(\bm{v})$ is a vector
comprised of the functions, $\psi(\bm{v})$, that generate the selected moments, $\bm{U}$.
The maximum order of the particle velocity, $\bm{v}$, that one includes into $\bm{m}$, defines the order of the method.
In fourth-order maximum-entropy methods, we consider terms up to $v^4$.
The maximum-entropy VDF of Eq.~\eqref{eq:max-ent-VDF-definition-exp} is positive by construction,
and its general shape permits one to reproduce a wide range of 
non-equilibrium distribution functions \cite{boccelli2023gallery}.
To summarize, the maximum-entropy procedure goes as follows:

\begin{enumerate}
  \item At a given time, $t$, and position, $\bm{x}$, one knows the value of the moments of interest, 
        $\bm{U}$;
  \item One then needs to find the parameters, $\bm{\alpha}$, such that the maximum-entropy distribution function of Eq.~\eqref{eq:max-ent-VDF-definition-exp} integrates to the moments, $\bm{U}$;
  \item The resulting VDF can be integrated in velocity space, giving the value of the closing fluxes.
\end{enumerate}

\noindent This allows one to march in time to the next time step.
To date, an analytical solution to the entropy maximization problem is only available for second-order methods.
In all other cases, including the fourth-order methods employed in this work, one needs to find the coefficients, $\bm{\alpha}$, numerically.
Since this needs to be done for every position in space and for each time step, this makes a solution of this problem extremely time-consuming.
Hardware acceleration was proposed as a solution to alleviate this issue \cite{schaerer2017efficient,garrett2015optimization}.
However, an even more convenient approach was developed in \cite{mcdonald2013affordable}, for a selected fourth-order maximum-entropy system, 
where an approximated solution to the entropy maximization problem was proposed.
This approximation gives the closing fluxes, $\bm{F}$, as an explicit function of the state vector, $\bm{F} = \bm{F}(\bm{U})$, and permits one to avoid
the expensive numerical iterations.
The computational cost is reduced by orders of magnitude, and numerical simulations become computationally affordable.
In the following, this is referred to as the approximated interpolative closure. 
This approach is not available for all maximum-entropy systems, but, to date, has been developed for the 14-moment \cite{mcdonald2013affordable} and the 21-moment  \cite{giroux2021approximation} systems, which are fourth-order members of the maximum-entropy family of moment methods.
The former is the focus of this work.
In Sections~\ref{sec:5-mom-system-1D1V} and \ref{sec:14-mom-system-3D3V}, we specify the moment systems that are solved in this work. 


\subsection{1D1V gas: the 5-moment system}\label{sec:5-mom-system-1D1V}

It is convenient to initially consider the simplified case of a gas possessing a single translational degree of freedom.
We refer to such a gas as one-dimensional in both space and velocity, or 1D1V.
The space coordinate is $\bm{x} \equiv x \, \hat{\bm{x}}$, while the particle velocity vector has a single component, 
or $\bm{v}\equiv v \hat{\bm{x}}$.
%
%
With respect to the 3D case, the moments of a 1D1V gas are obtained as single integrals over the (scalar) particle velocity, $v$.
The first six moments are
\begin{equation}\label{eq:1D1Vmoments-kinetic-def}
  \begin{cases}
  \rho = m \int f(v) \, \mathrm{d} v \, , \\
  \rho u = m \int v \, f(v) \, \mathrm{d} v \, , \\
  P = m \int (v - u)^2 \, f(v) \, \mathrm{d} v \, , \\
  q = m \int (v - u)^3 \, f(v) \, \mathrm{d} v \, , \\
  r = m \int (v - u)^4 \, f(v) \, \mathrm{d} v \, , \\
  s = m \int (v - u)^5 \, f(v) \, \mathrm{d} v \, ,
  \end{cases}
\end{equation}

\noindent where the hydrostatic pressure is simply $P \equiv P_{xx}$, and subscripts are dropped for simplicity.
As for the 3D case, we employ here a definition of the heat flux, $q$, that differs from 
the traditional fluid dynamic definition by a factor of $1/2$.
At equilibrium, $q^\mathrm{eq} =0$ and $r^\mathrm{eq} = 3 P^2/\rho$.
For a 1D1V gas, the fourth-order maximum-entropy method results in a system of five balance laws, in the form of Eq.~\ref{eq:PDEs-balance} (see  \cite{mcdonald2013affordable}),
\begin{subequations}\label{eq:ch-theory-5mom-sys-eqs}
  \begin{equation}
      \frac{\partial \rho}{\partial t} 
    + \frac{\partial}{\partial x} \left( \rho u \right) = 0 \, ,
  \end{equation}
  \begin{equation}
      \frac{\partial}{\partial t} \left( \rho u \right)
    + \frac{\partial}{\partial x} \left( \rho u^2 + P \right) = 0 \, ,
  \end{equation}
  \begin{equation}
      \frac{\partial}{\partial t} \left( \rho u^2 + P \right)
    + \frac{\partial}{\partial x} \left( \rho u^3 + 3 u P + q \right) = 0 \, ,
  \end{equation}
  \begin{equation}
      \frac{\partial}{\partial t} \left( \rho u^3 + 3 u P + q \right)
    + \frac{\partial}{\partial x} \left( \rho u^4 + 6 u^2 P + 4 u q + r \right) = - q/\tau \, ,
  \end{equation}
  \begin{equation}
      \frac{\partial}{\partial t} \left( \rho u^4 + 6 u^2 P + 4 u q + r \right)
    + \frac{\partial}{\partial x} \left( \rho u^5 + 10 u^3 P + 10 u^2 q  
      + 5 u r + s \right) = -\left( 4 u q + r - 3 P^2/\rho \right)/\tau \, .
  \end{equation}
\end{subequations}

\noindent The left-hand side terms originate from the streaming operator of the kinetic equation, while the 
right-hand side terms arise from the BGK collision operator and relax the moments to their Maxwellian value.
The fifth-order moment, $s$, appearing in the fluxes, is a closing moment, and its expression is discussed in Section~\ref{sec:phys-real-junk-closing-1D1V}.

\subsubsection{Physical realizability, Junk subspace and closing moment}\label{sec:phys-real-junk-closing-1D1V}

\begin{figure}[htpb]
  \centering
  \includegraphics[width=0.4\textwidth]{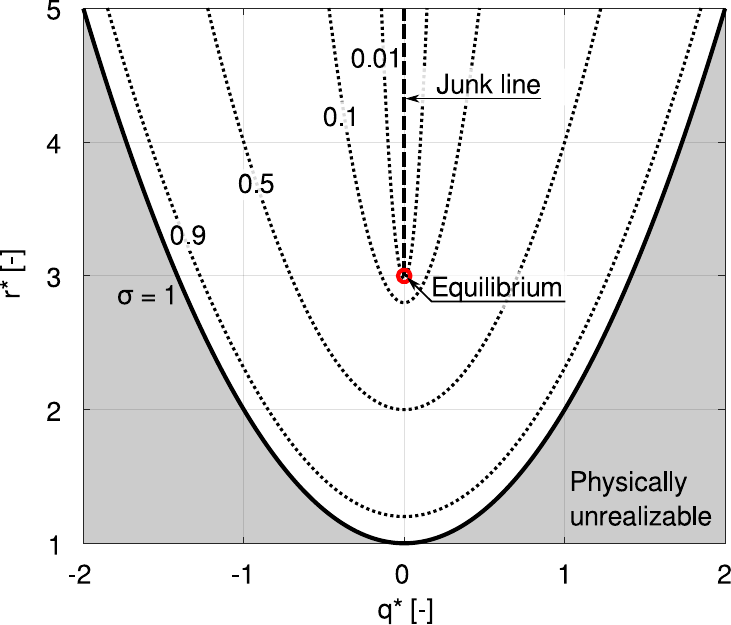}
  \caption{Physical realizability boundary, Junk subspace and equilibrium in dimensionless moment space, for the 1D1V 5-moment system.
           Dotted parabolas represent contours of the parabolic mapping parameter, $\sigma$.}
  \label{fig:momspace-realizability-5mom}
\end{figure}

After moving to a local rest frame, where the bulk velocity is $u=0$, and after non-dimensionalizing the system, 
the gas state can be described by two parameters, $q^\star$ and $r^\star$.
It can be shown that not all combinations of such states are physically possible.
Indeed, solving the Hamburger moment problem \cite{hamburger1944hermitian}, one can show that only states that satisfy
the inequality
\begin{equation}
  r^\star \ge q^{\star\, 2} + 1 
\end{equation}

\noindent can be realized by a non-negative distribution function.
The equality corresponds to a parabola in dimensionless moment space (see Fig.~\ref{fig:momspace-realizability-5mom}) and is 
referred to as the physical realizability boundary.
Also, Junk has shown \cite{junk2002maximum} that the fourth-order maximum-entropy method cannot reproduce a (zero-measure) set of states,
located on the line $q^\star = 0, r^\star \ge 3$.
This is shown in Fig.~\ref{fig:momspace-realizability-5mom} and is referred to as the Junk subspace or simply Junk line.
The impossibility to reproduce these states appears as a singularity in the convected fluxes:
as the gas state approaches the Junk line, the closing moment, $s$, goes to infinity \cite{groth2009towards,mcdonald2013towards}.
The wave speeds of the flux Jacobian also approach infinity, as the Junk line is approached.
This has important numerical consequences, as discussed throughout this work.

As discussed, to close the system, one needs an expression for the closing moment, $s$.
In this work, we employ the approximated interpolative closure by McDonald \& Torrilhon \cite{mcdonald2013affordable}.
First, one defines a parabolic mapping of the phase space, with a parameter 
\begin{equation}\label{eq:sigma-mapping-5mom}
  \sigma = \frac{1}{4} \left[ 3 - r^{\star} +\sqrt{(3 - r^\star)^2 + 8 q^{\star\, 2}} \right] \, .
\end{equation}

This definition is designed such that $\sigma = 1$ recovers the physical realizability boundary, 
while the Junk line is approached for $\sigma \to 0$.
The closing moment, $s$, is then defined as a function of $\sigma$ and $q^\star$ (the formula slightly differs from the original paper, 
as a power of $1/2$ was later shown to be more accurate),
\begin{equation}\label{eq:closure-1D1V-s-approx}
  s = \frac{q^{\star\, 3}}{\sigma^2} + (10 - 8 \sigma^{1/2})q^\star \, .
\end{equation}

\noindent During numerical calculations, the solution might approach the Junk line ($\sigma \to 0$), 
leading to extremely large values of the closing moment and of the wave speeds. 
As also discussed in \cite{mcdonald2013affordable}, this problem can be mitigated by performing an artificial limitation 
on $\sigma$:
one computes a limited value, $\bar{\sigma} = \mathrm{max}(\sigma, \sigma_\mathrm{lim})$, and uses it for computing 
the closing moment, $s$, and the wave speeds. 
A full discussion of the limiting procedure is given in \ref{sec:appendix-sigma-lim}.
Notice that a poor choice of $\sigma_\mathrm{lim}$ might affect the accuracy of the results, as discussed in Section~\ref{sec:sod-shock}.
From our observations, taking a value of $\sigma_\mathrm{lim}$ equal to $10^{-4}$ or $10^{-5}$ is generally a safe choice in terms of accuracy.
The value of $\sigma_\mathrm{lim}$ also influences the computational cost (Section~\ref{sec:sod-shock}).


\subsection{3D3V gas: the 14-moment system}\label{sec:14-mom-system-3D3V}

For a classical gas, whose particles have three degrees of freedom, one can formulate various fourth-order maximum-entropy 
systems \cite{levermore1996moment}. 
The simplest one is composed of 14 moments: the density, $\rho$, the bulk velocity, $u_i$, the pressure tensor, $P_{ij}$,
the heat flux vector, $q_i$ and the contracted fourth-order moment, $R_{iijj}$.
The governing equations for the 14 moments are
\begin{subequations}
\begin{equation}
    \tfrac{\partial}{\partial t} \rho 
    + \tfrac{\partial}{\partial x_i} \left( \rho u_i \right) = C_1  \, ,
\end{equation}
\begin{equation}
    \tfrac{\partial}{\partial t} \left( \rho u_i \right)
    + \tfrac{\partial}{\partial x_j} \left( \rho u_i u_j + P_{ij} \right) = C_{i} \, ,
\end{equation}
\begin{equation}
    \tfrac{\partial}{\partial t} \left( \rho u_i u_j + P_{ij} \right)
    + \tfrac{\partial}{\partial x_k} \left( \rho u_i u_j u_k + u_i P_{jk} + u_j P_{ik} 
     + u_k P_{ij}  + Q_{ijk} \right) = C_{ij} \, ,
\end{equation}
\begin{multline}
    \tfrac{\partial}{\partial t} \left( \rho u_i u_j u_j + u_i P_{jj} + 2 u_j P_{ij} + Q_{ijj} \right)  \\
    +\tfrac{\partial}{\partial x_k} \left(\rho u_i u_k u_j u_j  + u_i u_k P_{jj}
     + 2 u_i u_j P_{jk} + 2 u_j u_k P_{ij} + u_j u_j P_{ik}  \right. \\
    \left.  + u_i Q_{kjj} + u_k Q_{ijj} + 2 u_j Q_{ijk} + R_{ikjj} \right) = C_{ijj} \, ,
\end{multline}
\begin{multline}
    \tfrac{\partial}{\partial t} \left( \rho u_i u_i u_j u_j + 2 u_i u_i P_{jj} + 4 u_i u_j P_{ij} + 4 u_i Q_{ijj} + R_{iijj} \right) \\
    + \tfrac{\partial}{\partial x_k} \left( \rho u_k u_i u_i u_j u_j + 2 u_k u_i u_i P_{jj} + 4 u_i u_i u_j P_{jk}+ 4 u_i u_j u_k P_{ij}  + 2 u_i u_i Q_{kjj} \right. \\
    \left.  + 4 u_i u_k Q_{ijj} + 4 u_i u_j Q_{ijk} + 4 u_i R_{ikjj} + u_k R_{iijj} + S_{kiijj} \right) = C_{iijj} \, ,
\end{multline}
\end{subequations}

\noindent where repeated indices imply summation. 
To represent (twice) the heat flux vector, we often employ the shorthand $q_i \equiv Q_{ijj}$.
The right-hand side terms relax the moments towards their equilibrium value following the BGK collision operator,
\begin{equation}
  C_1 = 0 \  \ , \ \ \ C_i = 0 \ \ , \ \ \ C_{ij} = - (P_{ij} - P\delta_{ij})/\tau \, ,
\end{equation}

\noindent where $\delta_{ij}$ is the Kronecker delta, and the hydrostatic pressure is obtained from the trace of the pressure tensor, $P = P_{ii}/3$.
The other terms are
\begin{subequations}
\begin{equation}
  C_{ijj} = -(2 u_j P_{ij} + Q_{ijj} - 2 u_i P)/\tau  \ \ , \ \ \ 
\end{equation}
\begin{equation}
  C_{iijj} = -\left[\left( 4 u_i u_j P_{ij} + 4 u_i Q_{ijj} + R_{iijj} \right) - \left( 4 |u|^2 P + 15 P^2/\rho \right) \right]/\tau \, .
\end{equation}
\end{subequations}


\subsubsection{Physical realizability, Junk subspace and closing moments}

As discussed for the 1D1V gas, not all states are compatible with a non-negative distribution function.
In this 3D3V case, physical realizability for a gas modelled with the mentioned 14 moments requires that 
\begin{equation}\label{eq:phys-realizability-3D3Vgas}
  R_{iijj} \ge Q_{kii}(P^{-1})_{kl} Q_{ljj} + \frac{P_{ii} P_{jj}}{\rho} \, .
\end{equation}

\noindent Notice that, while for a one-dimensional gas this condition is both necessary and sufficient,
for a three-dimensional gas it is only sufficient.
The 14-moment maximum-entropy closure also embeds a singularity in the moment space (Junk subspace), that is defined by
\begin{equation}
  Q_{ijj} = 0 \ \ , \ \ \ R_{iijj} \ge \frac{2 P_{ji} P_{ij} + P_{ii} P_{jj}}{\rho} \, .
\end{equation}

\noindent The third-order tensor, $Q_{ijk}$, the second-order tensor, $R_{ikjj}$, and the vector, $S_{kiijj}$, that appear in the fluxes, need to be closed.
We follow the approximated interpolative closure of \cite{mcdonald2013affordable}:
first, a parabolic mapping of the moment space is performed, with a parameter
\begin{equation}\label{eq:sigma-mapping-14mom}
  \sigma = \frac{1}{4 P_{ij}P_{ji}} \left[ 2 P_{ij} P_{ji} + P_{ii}P_{jj} - \rho R_{iijj} \vphantom{\sqrt{P_k^2}}\right. \\
           \left. + \sqrt{ \left( 2 P_{ij} P_{ji} + P_{ii}P_{jj} - \rho R_{iijj}\right)^2 + 8 \rho P_{mn} P_{nm} Q_{kii} \left(P^{-1}\right)_{kl}Q_{ljj}   }\right] \, .
\end{equation}

\noindent This definition recovers the physical realizability boundary of Eq.~\eqref{eq:phys-realizability-3D3Vgas}
when $\sigma = 1$, and it approaches the Junk subspace for $\sigma \to 0$.
Since the Junk subspace is associated with a singularity of the fluxes, one numerically imposes a lower bound to the parabolic mapping parameter, 
$\sigma \ge \sigma_{\mathrm{lim}}$, often taking $\sigma_{\mathrm{lim}} = 10^{-4}$ or $10^{-5}$, as previously discussed for the one-dimensional gas.
The effect of this threshold is discussed throughout this work.
The closing moments are then approximated as follows.
The heat flux tensor is written as
\begin{equation}\label{eq:closure-14mom-Qijk-approx}
  Q_{ijk} = K_{ijkm} Q_{mnn} \, ,
\end{equation}

\noindent the tensor $R_{ijkk}$ is approximated as
\begin{equation}\label{eq:closure-14mom-Rijkk-approx}
  R_{ijkk} = \frac{1}{\sigma}Q_{ijl}(P^{-1})_{lm}Q_{mkk} + \frac{2 (1-\sigma) P_{ik} P_{kj} + P_{ij}P_{kk}}{\rho} \, ,
\end{equation}

\noindent and the vector $S_{ijjkk}$ is written as 
\begin{equation}\label{eq:closure-14mom-Sijjkk-approx}
  S_{ijjkk} = \frac{1}{\sigma^2} P_{kn}^{-1}P_{lm}^{-1} Q_{npp} Q_{mjj} Q_{ikl} 
+ 2 \sigma^{1/2} \frac{P_{jj} Q_{ikk}}{\rho} + (1 - \sigma^{1/2}) W_{im} Q_{mnn} \, ,
\end{equation}

\noindent where $\sigma^{1/2}$ is used, instead of the value $\sigma^{3/2}$ appearing in the original reference, since it was later shown to
be more accurate, and where some auxiliary matrices are defined as
\begin{subequations}
\begin{equation}
  B_{lm} = 2 P_{lm}(P^2)_{\alpha \alpha} + 4 (P^3)_{lm} \, ,
\end{equation}
\begin{equation}
  K_{ijkm} = \left[ 2P_{il}(P^2)_{jk} + 2 P_{kl} (P^2)_{ij} + 2 P_{jl}(P^2)_{ik} \right] B_{lm}^{-1} \, ,
\end{equation}
\begin{multline}
  W_{im} = \frac{1}{\rho} \left[ 2 P_{il}(P_{\alpha\alpha})^3 + 12P_{il}(P^3)_{\alpha\alpha} + 14 (P^2)_{\alpha\alpha} (P^2)_{il} + 20 P_{\alpha\alpha}(P^3)_{il} \right.\\ 
\left.+ 20(P^4)_{il} - 2(P^2)_{\alpha\alpha}P_{\beta\beta}P_{il} - 6(P_{\alpha\alpha})^2(P^2)_{il}\right]B_{lm}^{-1} \, .
\end{multline}
\end{subequations}

\noindent With the presented definitions, the system is closed and can be solved numerically.


\subsubsection{Considerations on the 14-moment VDF}

The discussed closure bypasses completely the computation of the maximum-entropy distribution function.
In this work, the coefficients $\alpha_i$ are not computed explicitly during numerical simulations, and this speeds-up dramatically the computations.
An explicit knowledge of these coefficients is unnecessary for the sake of solving the presented system.
Nonetheless, we detail here the shape of the 14-moment distribution function associated with the presented closure,
\begin{equation}
  f_{14} = \exp\left[ \alpha_0 + \alpha_i v_i + \alpha_{ij} v_i v_j + \alpha_{ijj} v_i v^2 + \alpha_{iijj} v^4 \right] \, .
\end{equation}

\noindent This VDF can be seen to reduce to a Maxwellian, when $\alpha_{ijj}=0$, $\alpha_{iijj}=0$ and when the elements $\alpha_{ij}$ 
form an isotropic and diagonal tensor.
In this special case, the value of the remaining non-zero coefficients is easily found.
In the other cases, one is in a non-equilibrium situation, and a simple relation between the 
coefficients and the 14 available moments is unknown, to date.

It should be noted that the three coefficients $\alpha_{ijj}$ multiply terms that are proportional to the 
cube of the velocity.
Therefore, these coefficients can be expected to be associated with asymmetries of the VDF, and to a non-zero heat flux.
On the other hand, the coefficient $\alpha_{iijj}$ allows one to reproduce a non-Maxwellian kurtosis,
that is the fourth-order moment of the VDF.
To fix the ideas, a situation of non-Maxwellian kurtosis happens whenever low-collisional streams of particles cross each-other.
Some examples are analyzed in Section~\ref{sec:sod-shock} (low-density Riemann problem) and in Section~\ref{sec:two-jets} (rarefied crossing jets). 
Alternatively, these situations may happen in the presence of chemical reactions, that selectively deplete the high-energy tails of the distribution function.

Another noteworthy scenario where one may need to employ a proper modelling of the non-Maxwellian 
kurtosis (and thus, a moment method analogous to the one presented here) is the simulation of 
hypersonic blunt-bodies flying in the rarefied portion of the atmosphere.
In such conditions, the wake is often characterized by an extremely low density, and the gas that 
fills the wake from the sides of the blunt body crosses at the center-line, resulting in a ring-like
VDF \cite{bariselli2018aerothermodynamic,boccelli2023modeling}.  

Finally, we remark that, whereas a direct knowledge of the VDF is not needed in the present work, there are cases where 
a full computation of the coefficients vector, $\bm{\alpha}$, might be required.
For instance, an accurate modelling of chemical reactions, as well as non-equilibrium collision frequencies, may require this,
unless non-equilibrium formulas and approximations are developed.
We suggest this as a future research endeavour.


\section{Numerical solution}\label{sec:numerical-solution}

In this work, the maximum-entropy system is solved numerically using a cell-centered finite-volume formulation \cite{leveque2002finite},
employing either a forward Euler or a midpoint Euler (2\textsuperscript{nd} order Runge-Kutta) explicit time marching schemes.
In the rarefied conditions considered in this work, the collisional source terms do not present a significant stiffness, and an explicit scheme is sufficient.
The maximum-entropy system is implemented and solved with the Hyper2D solver \cite{boccelli2023hyper2d}.

\subsection{Overview of the possible numerical flux functions}

For solving the fourth-order maximum-entropy system, it is highly advantageous to employ flux functions that do not require a detailled knowledge of the 
system eigenstructure. 
Indeed, an analytical expression for the eigenvalues (and eigenvectors) of the flux Jacobian is to date unknown.
In one physical dimension (5-moment system), one can easily compute them numerically (as done for instance in \cite{groth2009towards}).
Instead, in three dimensions, computing the eigenvalues of the $14\times14$ flux Jacobian, in every cell at every time step, would be a major computational effort.

As the full eigenstructure is not available, one needs an alternative to the commonly employed Roe or HLLE schemes, to cite a few \cite{roe1986characteristic,einfeldt1988godunov}.
Other successful methods commonly employed in fluid dynamics, such as the schemes of the AUSM family \cite{liou2010evolution}, are still to be formulated for the
maximum-entropy system.
As an alternative, one could employ kinetic fluxes \cite{pullin1980direct,perthame1992second}, that have the benefit of being consistent with kinetic theory.
This choice have been applied to maximum-entropy systems in the past \cite{garrett2015optimization}, however, it requires one to know the shape of the local VDF.
This is an information that comes at the cost of running the full entropy maximization procedure in every cell of the domain and at every time step, 
and we thus prefer to avoid it (opting instead for the interpolative closure).
The simple Lax-Friedrichs method has been proficiently employed in the past to solve maximum-entropy systems \cite{mcdonald2016approximate}, 
but it is overwhelmingly diffusive in highly rarefied and supersonic conditions, where the Junk line is approached closely,
and the fastest wave speeds become very large.
The HLL scheme has also been employed in previous maximum-entropy simulations \cite{tensuda2015application}, but for relatively high-collisional cases, and employing 
crude approximations for the eigenvalues, unsuitable for supersonic rarefied conditions.

In Section~\ref{sec:approx-wave-speeds}, we present an approximation for the maximum and minimum wave-speeds of the 14-moment system.
Our approximation is rough, but allows us to run supersonic simulations at any degree of rarefaction, using the Rusanov (local Lax-Friedrichs) 
scheme \cite{toro2013riemann}. 
This is the scheme of choice for this work.
It should be noted that our approximated wave speeds work well in conjunction with the HLL method, in low-Knudsen number conditions.
However, at large Knudsen numbers, the HLL scheme produces stable but large numerical oscillations.
We attribute this to the inaccuracies embedded in our wave speed approximation. 
Other possible schemes, to be investigated in a future work, include the 
Nessyahu-Tadmor scheme \cite{nessyahu1990non}, that was successfully applied to moment methods in the past \cite{au2001shock}, 
the Kurganov-Tadmor \cite{kurganov2000new} and the FORCE-$\alpha$ schemes \cite{toro2020low}.


\subsection{Second-order spacial accuracy}\label{sec:MUSCL-for-moments}

Second-order accuracy in space can be achieved via a linear reconstruction of the solution at the interface (MUSCL approach \cite{van1979towards}), with a total-variation diminishing (TVD) slope limiter to cope with discontinuities.
In this work, we employ the symmetric van Albada limiter \cite{toro2013riemann}.

Typically, the reconstruction procedure is more robust if performed on primitive variables, as opposed to conserved variables, especially when simulating high-speed flows \cite{josyula2015hypersonic}.
This can be explained with the fact that, while all conserved moments are proportional to the same power of the density, different moments depend on different powers of the velocity.
For instance, considering the Euler equations, the conserved variables are the density, momentum and energy:
these are of $0$\textsuperscript{th},
$1$\textsuperscript{st} and $2$\textsuperscript{nd} order in the velocity, respectively.
When performing a linear reconstruction on the momentum, $\rho u$, this affects the total energy, $\rho E = \rho u^2/2 + P/(\gamma - 1)$, to a larger extent than the momentum itself.
The linear reconstruction of momentum may increase the term proportional to $u^2$ quadratically, 
while the energy is only increased linearly.
This may ultimately result in negative pressures and temperatures.
This effect is even stronger in higher-order moment methods:
in the 14-moment system, conserved variables range from the 
$0$\textsuperscript{th} to the $4$\textsuperscript{th} order in the velocity, increasing this problem dramatically.
Therefore, performing the reconstruction in primitive variables is even more important.
For the 14-moment system, the primitive variables are $\bm{Q}_{14} = \left(\rho, u_i, P_{ij}, q_i, R_{iijj}\right)$.

Besides this, one should pay additional care in the way the slope-limiting procedure is performed.
Ideally, one should perform the limiting on the eigencomponents of the system \cite{leveque2002finite}.
However, for simplicity, one often limits the primitive variables directly \cite{hirsch1990numerical}.
For the 14-moment maximum-entropy system, and employing our approximated wave speeds, we observe that this procedure is not sufficiently robust in 
supersonic and rarefied cases.
From our numerical tests, we suggest that a robust strategy consists in employing the same value of the slope limiter for reconstructing 
all primitive variables. 
The limiter should be the most stringent one, among all primitive variables.
This formulation is also suggested in \cite{toro2013riemann}, and still results in second-order spacial accuracy in smooth regions, 
but is more cautious and limits \textit{all} variables, whenever a discontinuity in \textit{any} of the fourteen primitive variables is detected.


\subsection{Hardware acceleration with GPUs}

The maximum-entopy systems can be easily implemented and solved on both CPUs and GPUs (graphics processing units).
The two-dimensional simulations shown in this work are obtained on NVIDIA GPUs, with the CUDA version of the 
Hyper2D solver \cite{boccelli2023hyper2d}.
A Tesla K20X (launch year, 2012; 6 GB onboard memory; compute capability CC 3.5) and an A100 (year, 2020; 40 GB memory model; CC 8.0) are tested, separately.
he test cases shown in this work are run on a single GPU.
No domain decomposition techniques, multi-GPU or hybrid CPU-GPU computations are exploited at this time.
For all cases discussed in this paper the memory usage was well below the GPU capacity.

We report the following observations.
First, it is recommended to compute the closing moments in dimensionless form, and to re-dimensionalize them only after the computation is done.
This holds for both CPU and GPU computations.
Non-dimensionalizing the moments before the computation of the closing fluxes makes the inversion of the matrices, appearing in the closure formulas, more robust.
Scenarios where this might play a role include rarefied problems such as the expansion of a jet into vacuum (see Section~\ref{sec:two-jets}), 
the simulation of light-weight particles, such as electrons \cite{boccelli202014,romano20022d}, and plasmas in general 
\cite{boccelli202214}:
in such problems, the large numerical disparity between the low mass density and the higher order moments would
otherwise challenge the interpolative closure.
When computing the closing moments in non-dimensional form, we were able to obtain robust computations using both double and single precision.
Conversely, when computing the closing moments in dimensional form, double-precision computations were always necessary.

One further point to be considered concerns the complexity of the interpolative maximum-entropy closure.
In the simplest implementation, the computation of the closing moments requires to define a number of variables.
In the older GPU architectures, this might fill all available registers.
In CUDA, a simple solution consists in modifying the launch configuration, allocating less threads for each block, although this 
may result in a slight loss in performance.
This problem was observed in the Tesla K20X, while no issue was observed in the more modern A100.

Considering the said points, we observe the following performance, for double precision calculations, and a launch configuration of 
$8\times8$ threads per block.
With respect to a single-core CPU computation (AMD Ryzen 5, 5600G, year 2021), our CUDA implementation achieved a speed-up of 21 times on the 
10-year-old Tesla K20X, and a speed-up of 342 times on the A100.
It should be stressed that our CUDA implementation is rather basic, and does not employ any advanced memory management, 
nor GPU-specific features.
Significant improvements in the GPU efficiency are expected by addressing these factors.
For further discussion, the reader is referred to \cite{torres2011understanding,tabik2015demystifying,wang2020memory}.


\section{1D test case: rarefied Sod shock-tube problem}\label{sec:sod-shock}

The shock-structure prediction is perhaps the most studied test case for moment systems \cite{mentrelli2021shock,mentrelli2021shock3rd,torrilhon_struchtrup_2004,laplante2016comparison}.
In this section, we consider the full solution to the Sod shock-tube problem \cite{sod1978survey}.
Previous moment-method studies of the shock-tube problem include \cite{soga2001application,fan2020accelerating,koellermeier2020spline,mcdonald2013towards}.
In this problem, the initial state of the gas embeds a discontinuity, that divides two otherwise uniform left ($L$) and right ($R$) 
states.
In the fully collisional limit, the solution to this Riemann problem is known to be self-similar, 
with similarity variable $\xi = x/T$, where $x$ is the spacial location and $T$ is the solution time.
Here, we study the solution of this problem from the continuum regime to collisionless conditions, analyzing the effect of the collisionality.
First, in Section \ref{sec:1D-Sod-shock-kinetic}, we obtain a kinetic solution, solving the kinetic BGK equation, Eq.~\eqref{eq:boltzmann-bgk-eq}, for a gas with a single translational degree of freedom (1D1V gas).
Then, in Section \ref{sec:1D-Sod-shock-5moments}, we solve the maximum-entropy system of equations, that
for a 1D1V gas simplifies to the 5-moment system of Eq.~\eqref{eq:ch-theory-5mom-sys-eqs}.

Different collisionalities can be studied by varying the initial density of the gas:
lower densities reduce the collision frequency and increase the mean free path.
Here, we select a different approach: 
the left and right gas densities are the same for all simulations, and we consider instead different simulation times.
At small simulated times (considerably lower than the inverse of the collision frequency), the gas particles do not have 
sufficient time to collide, 
and the simulation is practically free-molecular.
As time passes, the solution approaches the continuum regime.
Usually, one describes the effect of collisionality by use of the Knudsen number, $\mathrm{Kn}=\lambda/D$, 
where $\lambda$ is the mean free path and $D$ is a reference length  \cite{ferziger1972mathematical}.
In the case of a flow past a solid body, the reference length is easily obtained.
However, in the present case, such choice is not trivial, since the solution structure spreads in space, as time passes.
Different definitions of the Knudsen number, based on local gradients \cite{josyula2011review}, could be employed.
Instead, here, we use the simple definition $\mathrm{Kn}=\lambda_L/D$, based on the left-state mean free path, 
$\lambda_L$, and the length of the computational domain, $D$, that is selected as to embed the full solution profile.
With this definition, a large Knudsen number means that the Riemann problem is at the very beginning of its evolution, as the mean free 
path is much larger than the whole simulated domain.
On the other hand, a small value of $\mathrm{Kn}$ indicates that many collisions happen through the whole profile, and we can expect 
the traditional shock/contact-discontinuity/rarefaction-fan structure to emerge.
Notice that our definition of the Knudsen number is rather conservative, being based on the largest characteristic length, and on the highest density state.

The left and right initial states are taken at thermodynamic equilibrium.
The densities are fixed to atmospheric-like conditions,
$\rho_{L0} = 4~\si{kg/m^3}$ and $\rho_{R0}=1~\si{kg/m^3}$, while the initial bulk velocity is zero and the temperature
is $T_{L0}=T_{R0}=480~\si{K}$.
We simulate argon gas, assuming a constant elastic cross-section $\sigma=5.463\times10^{-19}~\si{m^2}$, 
and we model the effect of collisions through the BGK collision operator, as discussed in Eq.~\eqref{eq:boltzmann-bgk-eq}.
The mean collision time, $\tau$, is computed with the approximated formula, $\tau=(n \sigma v^{\mathrm{th}})^{-1}$, where $n$ is the local 
number density.
The thermal velocity is computed as $v^{\mathrm{th}}=\sqrt{8 k_B T/\pi m}$, with $k_B$ the Boltzmann constant and $m$ the mass of gas particles.
Notice that the expressions employed here for $\tau$ and $v^{\mathrm{th}}$ are not entirely consistent (or accurate) 
for a one-dimensional non-equilibrium gas.
This does not matter for the sake of the present work, where we only aim at comparing the different models, for analogous collisionalities.


\subsection{Kinetic solution}\label{sec:1D-Sod-shock-kinetic}

\begin{figure}[htb]
  \centering
  \includegraphics[width=0.9\columnwidth]{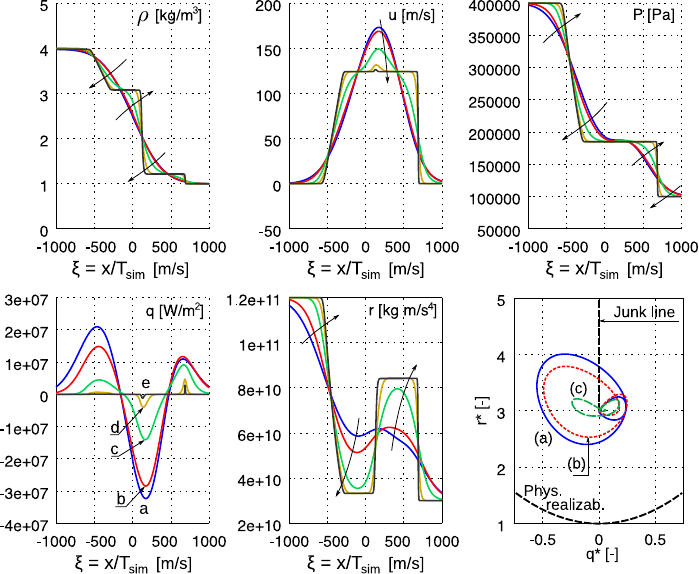}
  \caption{Moments obtained from the numerical solution of the BGK kinetic equation. Cases (a) to (e) are obtained at increasing simulated times, 
           that progressively approach local thermodynamic equilibrium.
           Arrows show the evolution of the moments with the increasing simulated time.
           The bottom-right panel shows the solution in the dimensionless moment space: 
           the Junk line is shown only for reference and does not affect anyhow the kinetic solution.
           The physical realizability condition is automatically satisfied by the kinetic solution.}
  \label{fig:kin-sol-sod-shock}
\end{figure}

The kinetic solution is obtained by a direct discretization of the 1D1V phase space \cite{mieussens2000discrete}, marching in time with 
an explicit integrator. 
Our numerical implementation is based on the 1D1V kinetic-theory version of the Hyper2D solver \cite{boccelli2023hyper2d}.
After computing the solution, its moments can be computed by direct integration of the distribution function,
following Eqs.~\eqref{eq:1D1Vmoments-kinetic-def}. 
Figure~\ref{fig:kin-sol-sod-shock} shows the moments of the kinetic solution for five different simulated times, 
$T_{\mathrm{sim}}$, equivalent to the following Knudsen numbers:
\begin{description}
  \item[a)] $\mathrm{Kn}\approx1$, $T_{\mathrm{sim}}=10^{-11}~\si{s}$;
  \item[b)] $\mathrm{Kn}\approx0.1$, $T_{\mathrm{sim}}=10^{-10}~\si{s}$;
  \item[c)] $\mathrm{Kn}\approx0.01$, $T_{\mathrm{sim}}=10^{-9}~\si{s}$;
  \item[d)] $\mathrm{Kn}\approx0.001$, $T_{\mathrm{sim}}=10^{-8}~\si{s}$;
  \item[e)] $\mathrm{Kn}\approx0.0001$, $T_{\mathrm{sim}}=10^{-7}~\si{s}$.
\end{description}


The small value of the simulation times (and thus of the space scales) are a result of the initial choice of the left and 
right densities, that is arbitrary.
At the very beginning of the simulation, for a simulated time much smaller than the mean collision time, $T_\mathrm{sim}\ll\tau$, 
the solution is low-collisional:
at such short time scales, the particle of the left and right states mix with a negligible probability of colliding.
In this case, the solution is easily obtained analytically and the moment profiles are completely continuous.
As the simulated time becomes comparable with the mean collision time, particles start to interact with each other.
Case a) shows this condition, where the simulated domain size is roughly equal to one mean free path.
In these conditions, the density and velocity profiles are still continuous and show no jumps.
As the simulated time increases, the expected shock/contact-discontinuity/rarefaction fan start to emerge and the profiles become self-similar.
Figure~\ref{fig:kin-sol-sod-shock} shows that the low-collisional simulations are associated with a much higher value of the heat flux.
Indeed, in such conditions the VDF is far from the equilibrium Maxwellian.
In moment space (Fig.~\ref{fig:kin-sol-sod-shock}-Bottom-Right), the solution is seen to deviate strongly from equilibrium.
It is interesting to observe that, in order to follow the very same path, the maximum-entropy system would need to cross 
the Junk subspace.


\subsection{5-moment maximum-entropy solution}\label{sec:1D-Sod-shock-5moments}

Traditional continuum models, such as the Euler equations of gas dynamics, are not able to reproduce the smooth 
profiles that characterize the collisionless situations of Section~\ref{sec:1D-Sod-shock-kinetic}.
In particular, the Euler equations always reproduces a combination of rarefaction fans, contact discontinuities and shock
waves, whatever the collisionality.
When solving the full Sod shock-tube problem with a moment method, one obtains a richer set of waves and smooth profiles, 
due to the more complicated eigenstructure of such methods.
Such waves are particularly evident in collisionless conditions, and collapse to the traditional 
shock/contact-discontinuity/rarefaction-fan in the continuum regime.

\begin{figure}[htb]
  \centering
  \includegraphics[width=1.0\columnwidth]{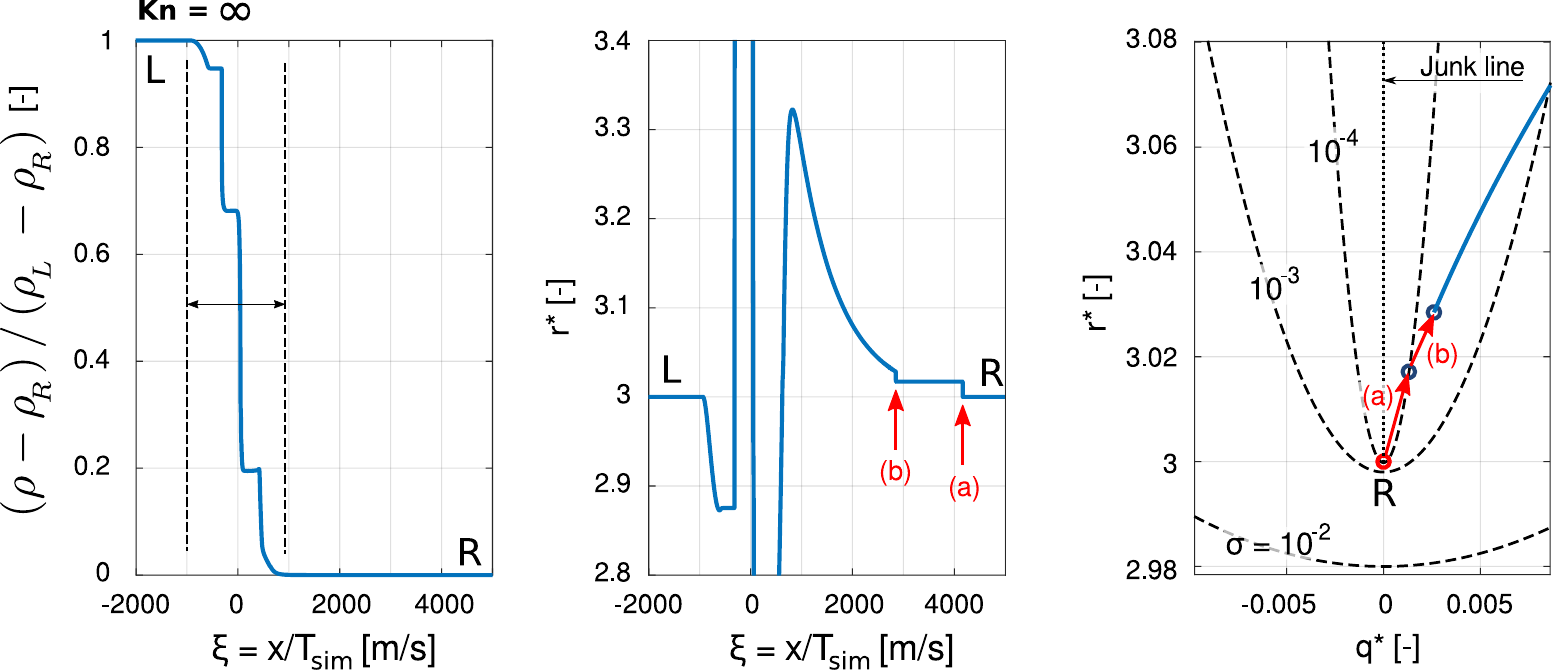}
  \caption{Solution of the 5-moment system for the Sod shock-tube problem test case, in collisionless conditions, 
           for $\sigma_\mathrm{lim}=10^{-4}$. 
           Left: normalized density profile; 
           the vertical dashed lines in the Left plot denote the main region, where the initial discontinuity breaks into various waves. 
           Centre: dimensionless fourth-order moment; a much faster wave can be observed ($a$), followed by a further discontinuity ($b$);
           Right: trajectory of the system in moment space, and effect of the discontinuities.
           The dashed lines indicate contours at constant $\sigma$.}
  \label{fig:5mom-fast-wave}
\end{figure}

Let us first consider collisionless conditions, or $\mathrm{Kn}\to\infty$.
The solution of the 5-moment system for this case is shown in Fig.~\ref{fig:5mom-fast-wave}.
A value of $\sigma_\mathrm{lim}=10^{-4}$ is employed in the simulations; for more details, see Section~\ref{sec:sigma-lim-and-computational-cost}.
From the density profile (Fig.~\ref{fig:5mom-fast-wave}-Left), the initial discontinuity can be seen to gradually break 
into a number of waves.
However, an analysis of the fourth-order moment (Fig.~\ref{fig:5mom-fast-wave}-Centre) reveals the presence of 
a much faster wave.
This wave is followed by a further discontinuity, that evolves as a continuous profile.
By inspecting the numerical results, the fast wave can be seen to affect all moments. 
For the fourth-order moment, this wave is clearly visible and its magnitude is the largest, 
while for lower-order moments, one needs to employ some magnification, or a logarithmic scaling.

\begin{figure}[htb]
  \centering
  \includegraphics[width=1.0\columnwidth]{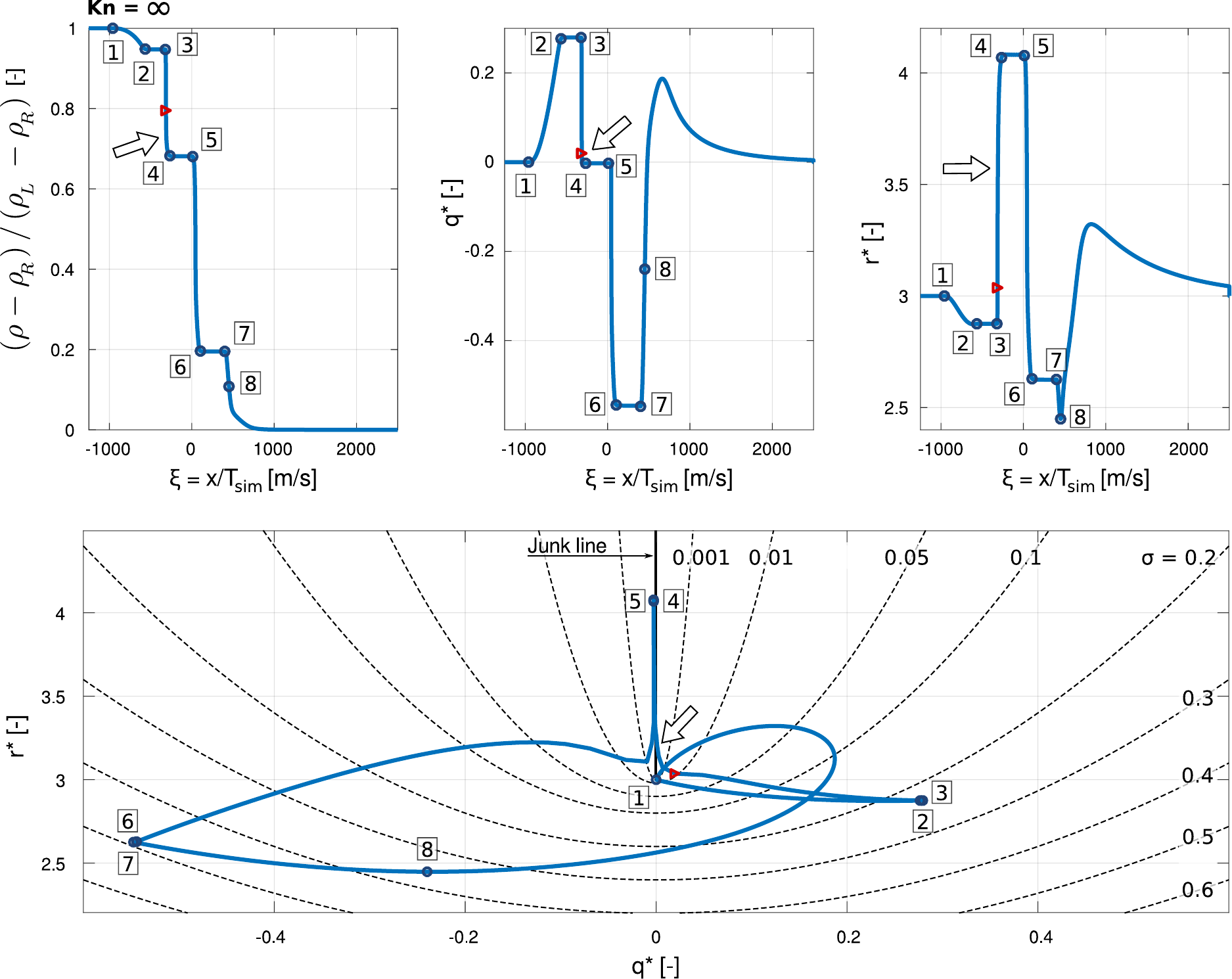}
  \caption{Solution of the 5-moment system for the Sod shock-tube problem test case, in collisionless conditions,
           for $\sigma_\mathrm{lim}=10^{-4}$. The top panels represent the scaled density, and dimensionless heat flux 
           and fourth-order moment. The bottom panel represents the solution in dimensionless moment space, superimposed to 
           lines at constant $\sigma$.  
           The point where the numerical solution crosses the Junk line is indicated with a white arrow. 
           A red marker is also employed to help mapping the crossing region among the different plots.}
  \label{fig:sod-5mom-structure-details-rho-q-r-momspace}
\end{figure}

Figure~\ref{fig:5mom-fast-wave}-Right shows the trajectory of the solution in dimensionless moment space, and allows
for an interpretation:
starting from the right state, $R$, the fastest wave brings the gas, from equilibrium, upwards in moment space, towards larger values of $r^\star$ 
(indicated by $(a)$ in Fig.~\ref{fig:5mom-fast-wave}-Right).
In the figure, the trajectory is not exactly vertical, due to the artificial limiting of the parabolic mapping parameter, 
here $\sigma_\mathrm{lim}=10^{-4}$.
The second discontinuity then pushes the system away from the Junk line.
After this, the system evolves into a continuous profile, until the next discontinuity is approached.
By testing different numerical values for the limiting parameter, $\sigma_\mathrm{lim}$, we observe that
\begin{itemize}
  \item The trajectory in moment space becomes vertical, as $\sigma_\mathrm{lim}$ is lowered;
  \item The amplitude of the discontinuity becomes smaller, for lower values of $\sigma_\mathrm{lim}$.
  \item The velocity of the discontinuity increases for a lower $\sigma_\mathrm{lim}$, forcing one to employ a smaller time step.
\end{itemize}

This last point is not surprising, since this wave is associated with the fastest eigenvalue of the system, 
that is known to diverge, as the Junk singularity is approached.
This discontinuity does not have an analog in the continuum limit,
since collisions would keep the system near equilibrium, preventing it from developing fast wave speeds.
In Figure~\ref{fig:sod-5mom-structure-details-rho-q-r-momspace}, we consider the remaining wave structure.
The simulations are run for a slightly longer time, such that the fast wave leaves the domain, and the slower waves develop across the numerical grid.
Starting from the left, the solution departs from equilibrium through a rarefaction-fan-like structure (points $1$--$2$).
Considering the plot in moment space, this rarefaction fan causes a non-zero heat flux, $q^\star$, 
and brings the fourth-order dimensionless moment, $r^\star$, to sub-Maxwellian values (the equilibrium value is $r^\star = 3$).
This rarefaction wave is followed by a plateau ($2$--$3$) and by a discontinuity ($3$--$4$).
Along this discontinuity, the solution appears to approach equilibrium again, but instead of reaching it, equilibrium is passed from the 
top, and the Junk line is crossed.
This is indicated by a large arrow in the figure.
A red triangular marker is also employed, as a reference to help identify this point.
It should be noted that this region develops in very few grid cells, and resolving it in moment space is challenging.
The plateau between points $4$ and $5$ is located at the left of the Junk line, but in extreme vicinity. 
For this state, the system wave speeds are expected to be very large.
Therefore, this region poses limits in the maximum allowed time step of the simulation.
After point $5$, in a few grid cells, the solution moves away from the Junk line and reaches another plateau ($6$--$7$).
The solution then evolves smoothly towards equilibrium ($7$--$8$ and forward), 
in a continuous structure also analyzed in Fig.~\ref{fig:5mom-fast-wave}.
This structure would terminate in the previously discussed fast waves, that are not shown in this plot.

\begin{figure}[htb]
  \centering
  \includegraphics[width=0.9\columnwidth]{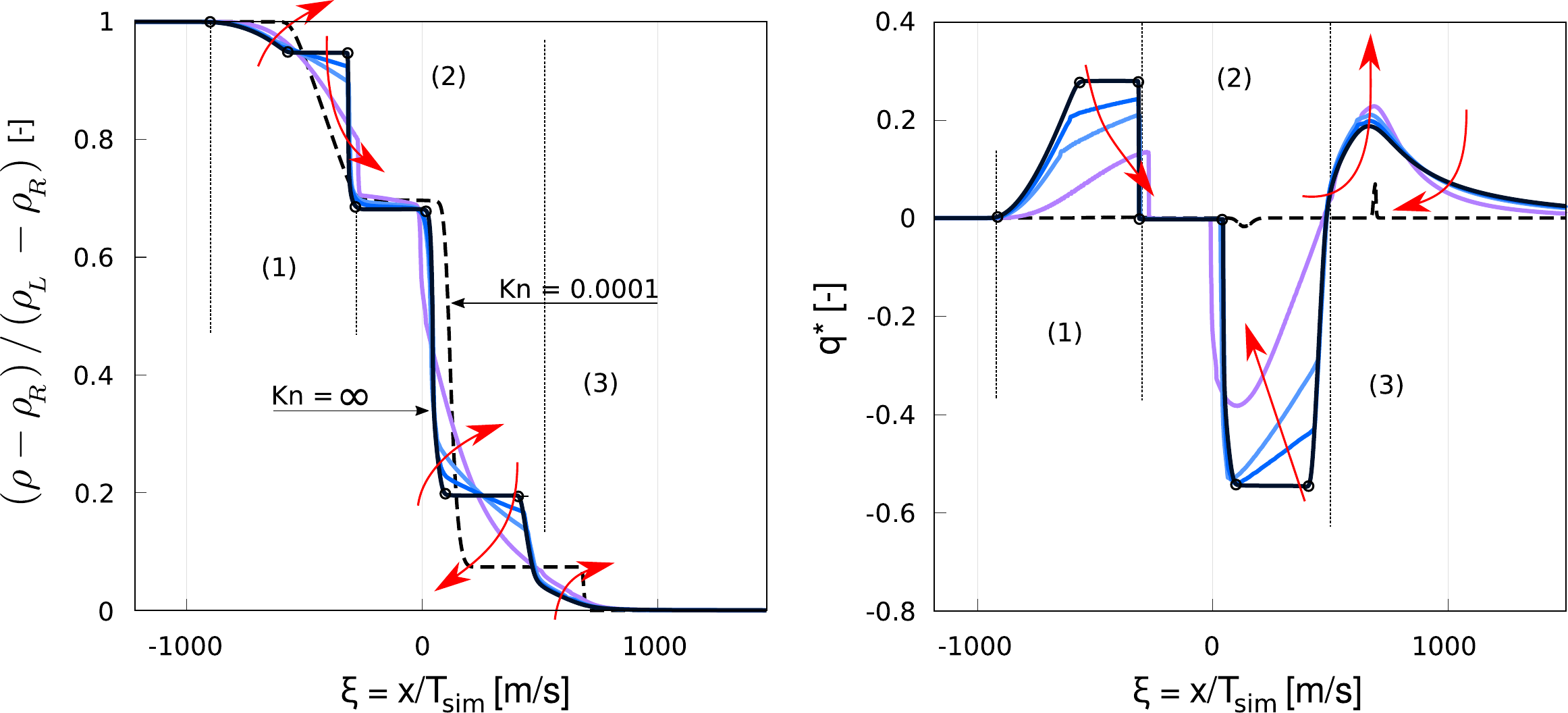}
  \caption{Solution of the 5-moment system for the Sod shock-tube problem.
           The red arrows indicate the evolution of the solution for increasingly longer simulated times,
           equivalent to the Knudsen numbers: $\mathrm{Kn}=\infty, 0.25, 0.1, 0.02$.  
           The dashed line is obtained for $\mathrm{Kn} = 0.0001$. 
           Regions (1), (2) and (3) indicate which parts of the initial wave structure evolve into the rarefaction fan, 
           contact discontinuity and shock wave respectively.}
  \label{fig:sod-5mom-structure-evolution}
\end{figure}

Under the effect of collisions, the wave structure of Fig.~\ref{fig:sod-5mom-structure-details-rho-q-r-momspace} is progressively modified,
and eventually recovers the continuum solution, characterized by a rarefaction fan, a contact discontinuity and
a traditional (discontinuous) shock wave.
The effect of collisions on the solution is shown in Fig.~\ref{fig:sod-5mom-structure-evolution}.
Different solutions are obtained at increased simulation times (equivalent to decreasing Knudsen number) and are superimposed on the same plot
by using the variable $\xi=x/T$.
In Fig.~\ref{fig:sod-5mom-structure-evolution}, three approximate regions are identified, to aid the interpretation.
The collisionless solution of Region (1) includes the rarefaction fan and the first discontinuity. 
As collisions equilibrate the solution, the rarefaction fan appearing in the collisionless case and the subsequent plateau appear 
to merge and evolve into the classical continuum rarefaction fan. 
Instead, the continuum contact discontinuity arises from the jumps located within Region (2). 
This appears clearly from the heat-flux plot (Fig.~\ref{fig:sod-5mom-structure-evolution}-Right), where the region of negative heat flux
gradually shrinks, until it appears as a small bump, in the collisional limit, located in correspondence with the contact discontinuity.
Finally, the smooth structure in Region (3) becomes compact, as collisionality increases, and eventually collapses into a shock wave.

\begin{figure}[htpb!]
  \centering
  \includegraphics[width=0.8\columnwidth]{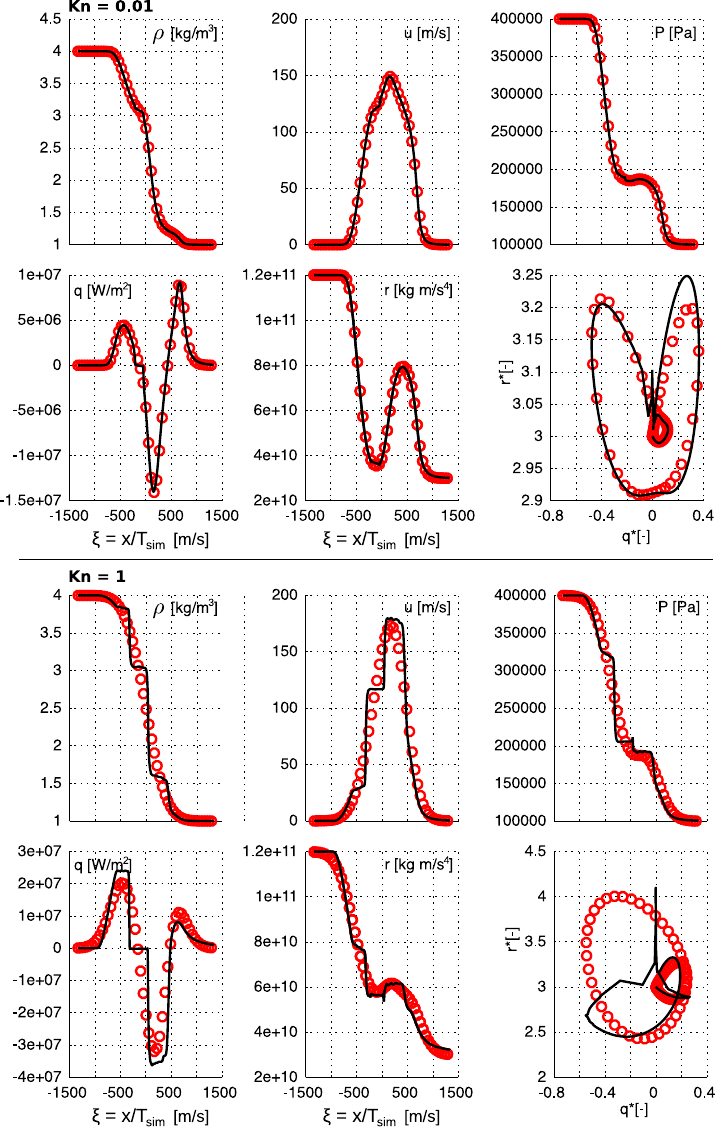}
  \caption{Solution of the Sod shock-tube problem for $\mathrm{Kn} = 0.01$ and $\mathrm{Kn} = 1$, based on the length of the simulation domain and on the mean free path of the left state.
           Solid lines: 5-moment system. Symbols: kinetic solution.}
  \label{fig:sod-5mom-vs-kinetic}
\end{figure}

In Figure~\ref{fig:sod-5mom-vs-kinetic}, we compare the 5-moment solution to 
the kinetic solution, for two selected Knudsen numbers.
At $\mathrm{Kn}=0.01$ (Fig.~\ref{fig:sod-5mom-vs-kinetic}-Top), the 5-moment solution is very close to the 
kinetic solution.
Instead, at $\mathrm{Kn}=1$ (Fig.~\ref{fig:sod-5mom-vs-kinetic}-Bottom), where the solution is globally 
farther from equilibrium, as expected, we observe a lower accuracy for the maximum-entropy method.
Indeed, the 5-moment system predicts a number of discontinuities that are not present in the kinetic solution.
Nonetheless, the full solution of the shock-tube problem appears to be reproduced to a reasonable accuracy. 
The overall profile is followed reasonably, and the location and amplitude of the peaks is well reproduced.
The moment-space plots in Fig.~\ref{fig:sod-5mom-vs-kinetic} show how the dynamics of the system happens farther from
equilibrium in the low-collisional case. This is compatible with the purely kinetic results shown in Fig.~\ref{fig:kin-sol-sod-shock}. 
At $\mathrm{Kn}=1$ (and even more so at larger Knudsen numbers), one can appreciate the effect of the Junk line, that is crossed seamlessly by the kinetic solution, 
but hampers the 14-moment solution.
The Junk line also affects the $\mathrm{Kn}=0.01$ simulation, although to a lesser extent.
Finally, it is important to stress that our definition of the Knudsen number is rather conservative, as we decided to consider
the whole domain as a reference length for the problem.
For additional discussions of the moment-space trajectory of shock waves, we refer the reader to \cite{groth2009towards,mcdonald2013towards}.


\subsection{Effect of $\sigma_\mathrm{lim}$ and computational cost}\label{sec:sigma-lim-and-computational-cost}

\begin{figure}[htpb!]
  \centering
  \includegraphics[width=1.0\columnwidth]{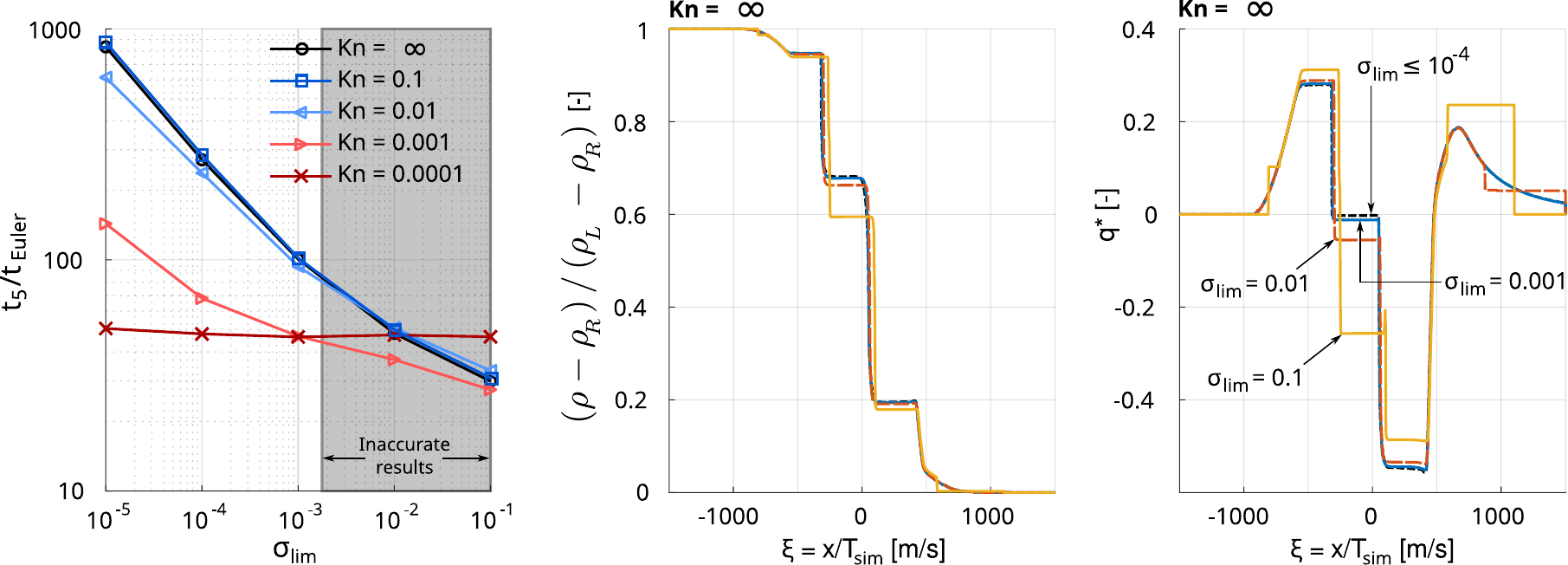}
  \caption{Effect of the limiting parameter, $\sigma_\mathrm{lim}$. Left: effect on the computational cost, for different Knudsen numbers. Centre and Right: accuracy of the computations and numerical artifacts, for the collisionless case. The correct solution is obtained for $\sigma_{\mathrm{lim}} \lesssim 10^{-4}$, and is shown as a black dashed line. A value of $\sigma_{\mathrm{lim}} = 10^{-3}$ also appears to be qualitatively acceptable.}
  \label{fig:sod-5mom-effect-of-sigma}
\end{figure}

In fourth-order maximum-entropy systems, the singularity in the closing fluxes causes the maximum wave speed to diverge, as the solution approaches the Junk line.
In the low-collisional Sod shock-tube problem, this happens in multiple regions of the domain.
In \cite{mcdonald2013affordable}, a limiting parameter, $\sigma_\mathrm{lim}$, was introduced in the interpolative closure, 
in order to artificially prevent the system from getting too close to the singularity.
Consequently, $\sigma_\mathrm{lim}$ also limits the maximum wave speed of the system, and has a direct effect on the computational cost
of time-explicit numerical simulations, through the CFL condition.
In Figure~\ref{fig:sod-5mom-effect-of-sigma}-Left, we analyze the computational cost associated with the solution of the 5-moment system, for the Sod shock-tube problem.
The analysis is performed for various values of the Knudsen number.
The computational cost is compared to the cost associated with the solution of the Euler equations of gas dynamics.
This comparison allows us (i) to obtain implementation-independent figures, and (ii) to gain an understanding on the affordability of the maximum-entropy method.
As the Euler equations are not able to reproduce non-equilibrium solutions, this is to be interpreted uniquely as a numerical comparison.
Overall, the 5-moment system appears to be from 30 to 1000 times more expensive than the Euler equations, as discussed in the following.

For sufficiently rarefied conditions (here, for $\mathrm{Kn} \ge 0.1$), the computational cost is approximately independent from the collisionality.
In this regime, non-equilibrium is very large, meaning that the solution explores a wide portion of moment space.
Below $\mathrm{Kn} = 0.01$, collisions start to limit appreciably the non-equilibrium; the system realizes lower wave speeds and the computational cost decreases. 
At $\mathrm{Kn} = 0.001$, the computational cost is much lower, as one approaches the continuum regime.
For lower Knudsen numbers, collisions dominate the problem:
at roughly $\mathrm{Kn} = 0.0001$, the BGK source terms impose the most strict timestep constraint.
In time-explicit integrators, this translates in a progressively higher computational cost.
From this point, it is therefore convenient to employ implicit time marching schemes (a simple point-implicit treatment of the source terms is sufficient).

Figure~\ref{fig:sod-5mom-effect-of-sigma}-Left shows that larger values of $\sigma_\mathrm{lim}$ are associated with lower computational costs, since the maximum wave speed is artificially limited.
However, this comes at the price of reducing the accuracy of the closure.
As shown in Fig.~\ref{fig:sod-5mom-effect-of-sigma}-Centre and Right (for the collisionless case), values of $\sigma_{\mathrm{lim}} \ge 0.01$ heavily modify the solution,
and introduce numerical artifacts.
For $\sigma_{\mathrm{lim}} = 0.001$, the solutions are not exact, but reasonable, and this value may be employed, if necessary, with proper care.
Instead, choices of $\sigma_\mathrm{lim} \le 0.0001$ are accurate.

Since the Knudsen number definition employed here is rather conservative, we expect 
the 5-moment method to behave slightly better than indicated, in real circumstances.
Also notice that the computational cost and the acceptable choices of $\sigma_\mathrm{lim}$ depend on the specific trajectory of the solution in moment space, 
and are expected to differ on a case-by-case basis.
For further discussion on the effect of the singularity on the 5-moment system, the reader is referred to \cite{schaerer2015singular}.

\section{Approximated wave speeds}\label{sec:approx-wave-speeds}

The wave speeds of the system are needed, during numerical simulations, for estimating the Courant number
and for most numerical flux formulations.
However, an exact closed-form expression for the wave speeds of fourth-order maximum-entropy systems is not available, to date, due to the
complexity of the entropy maximization procedure.
This implies that, in order to perform numerical simulations, one needs to build explicitly the flux Jacobian matrix, and then
to compute its eigenvalues, numerically.
For a one-dimensional gas (5-moment system), the increase in computational cost associated with this operation is typically manageable. 
On the other hand, for a three-dimensional gas, the flux Jacobian is a $14\times14$ matrix, whose expression is not trivial.
The computation of the eigenvalues for such a matrix easily becomes the most time-consuming part of a numerical simulation.
Therefore, an approximation is in order.

In the literature, in order to avoid the computational cost associated with this operation, 
the maximum and minimum wave speeds, $\lambda_{\mathrm{min}}^{\mathrm{max}}$, of the maximum-entropy system have often been assumed to be 
proportional to the local speed of sound, $a$, inflated by some arbitrary constant, $k$:
\begin{equation}\label{eq:wavespeed-u-pm-ka}
  \lambda_{\mathrm{min}}^{\mathrm{max}} = u \pm k a \, .
\end{equation}

\begin{figure}[htb]
  \centering
  \includegraphics[width=1.0\columnwidth]{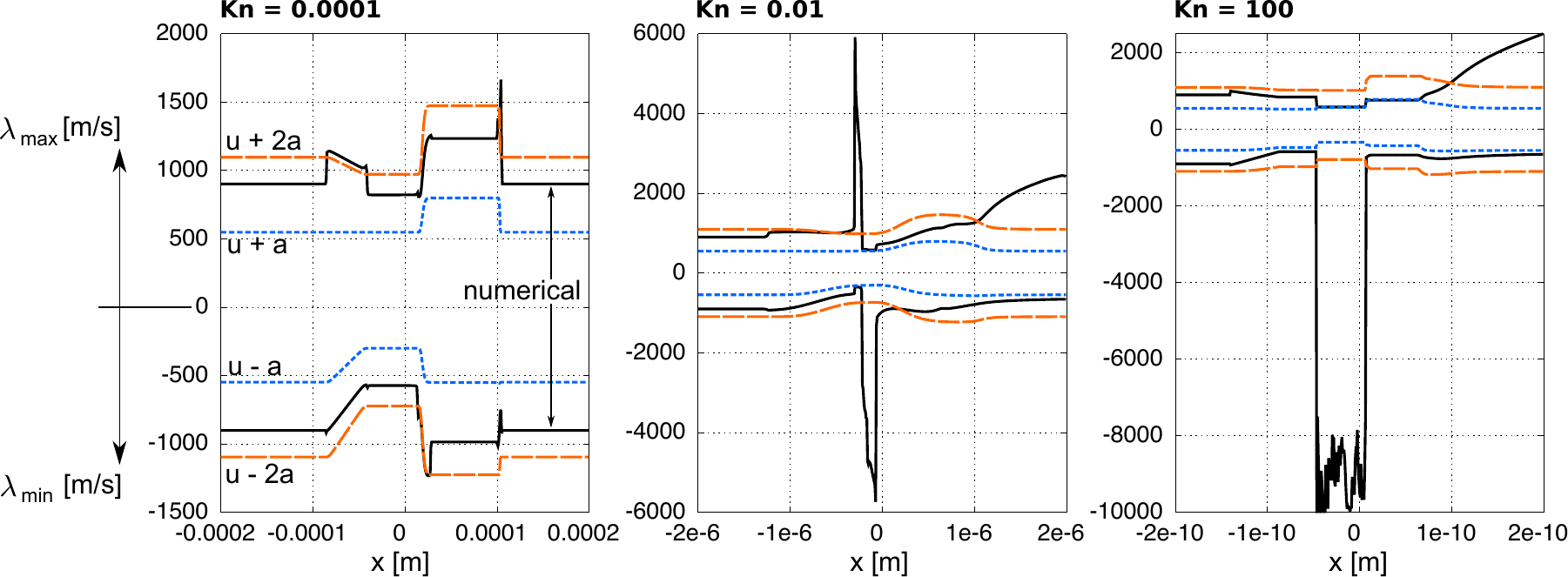}
  \caption{Maximum and minimum wave speeds of the 5-moment system for the Sod shock-tube problem. Solid black line: numerical wave speeds. Dashed lines: approximations from Eq.~\eqref{eq:wavespeed-u-pm-ka} with $k=1$ and $k=2$.}
  \label{fig:ws-5mom}
\end{figure}

\noindent The rationale behind this approximation is that, if $k$ is large enough, the simulation should remain stable.
This approach has the benefit of being simple, and was shown to be profitable in a number of conditions 
(see for instance \cite{tensuda2016multi}).
However, this approximation is not suitable for the present work, where the high rarefaction and supersonic speeds 
let the system depart significantly from equilibrium and also approach the Junk subspace.
Figure~\ref{fig:ws-5mom} compares this approximation to the actual numerical wave speeds of the solution.
We solve the 5-moment system for the conditions of the Sod shock-tube problem, discussed in Section~\ref{sec:sod-shock}, 
at different Knudsen numbers.
First, a solution of the Sod shock-tube problem was obtained using the numerical wave speeds; 
then, the approximated wave speeds of Eq.~\eqref{eq:wavespeed-u-pm-ka} were computed and plotted, a posteriori, based on such solution.

\begin{figure}[htb]
  \centering
  \includegraphics[width=1.0\columnwidth]{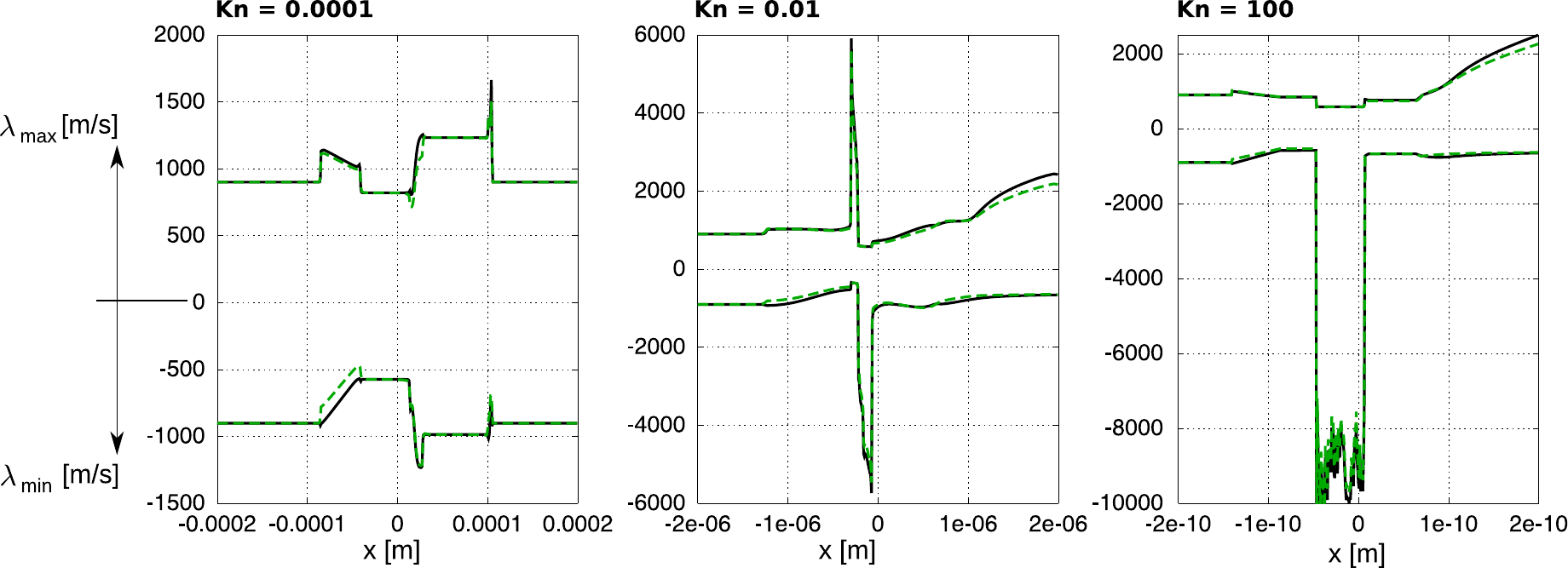}
  \caption{Maximum and minimum wave speeds of the 5-moment system for the Sod shock-tube problem. Solid black line: numerical wave speeds. Green dashed line: approximation from \cite{baradaran2015development}.}
  \label{fig:ws-5mom-baradaran}
\end{figure}

For continuum conditions ($\mathrm{Kn}\approx 0.0001$, in Figure~\ref{fig:ws-5mom}), the approximation of Eq.~\eqref{eq:wavespeed-u-pm-ka} appears reasonable.
However, as the rarefaction increases, the approximation quickly loses validity and largely underestimates the actual wave speeds, unless 
very large values of $k$ are employed. 
Most of the discrepancy appears around the contact discontinuity, where the maximum-entropy solution approaches the Junk line.
Taking a very large value of $k$ would thus result in a very large (and unnecessary) diffusion in most of the domain.
This justifies the need for better approximations.
For the one-dimensional 5-moment system, more accurate approximations for the wave speeds have been previously developed \cite{baradaran2015development}.
Figure~\ref{fig:ws-5mom-baradaran} shows these approximations as applied to the same Sod shock-tube problem, and confirms their accuracy, in most of the domain.
From our numerical investigations we noticed that, when applied to strongly-rarefied conditions, these approximations produce some oscillations if the 
HLL numerical flux scheme is employed. 
The Rusanov scheme was instead shown to be sufficiently dissipative to damp these oscillations, and we thus recommended it for rarefied conditions.

\subsection{Approximated wave speeds for the 14-moment system}\label{sec:ws-14mom-approx-formulas}

In this section, we propose an approximated expression for the maximum and minimum wave speeds of the 14-moment system, obtained 
by empirically extending the 5-moment approximations of \cite{baradaran2015development}.
In particular, the proposed wave speeds include the effect of the pressure anisotropy and of the three-dimensional heat flux.
These formulas are based on a number of fits in moment space.
Only the main points and the final results are reported here, while a more detailled discussion is given in 
\ref{appendix:approx-wave-speeds}.

Considering the wave speeds associated with the $x$ convective fluxes, the wave speed computation goes as follows.
First, one fixes the reference frame as to have a zero bulk velocity. 
In this frame, the gas state is fully defined by 11 independent moments, 
$(\rho,P_{ij},q_i,\sigma)$, where we have decided to employ the parabolic mapping parameter 
$\sigma$, in place of $R_{iijj}$, in virtue of Eq.~\eqref{eq:sigma-mapping-14mom}, and where $q_i=Q_{ijj}$.
The resulting wave speeds only depend on the thermodynamic state, and do not include the bulk velocity.

Our aim is now to simplify the picture, by introducing some further assumptions and simplifications.
By rotating the reference frame about $x$, the $x$-wave speeds do not change. 
We denote this new frame with a tilde symbol, and select the rotation angle such that it aligns the $y$ axis with the 
perpendicular component of the heat flux. 
In this new reference system, we have 
\begin{equation} 
    \tilde{q}_x=q_x \ \ , \ \ \ \tilde{q}_y = \sqrt{q_y^2 + q_z^2} \ \ , \ \ \  \tilde{q}_z=0 \, .
\end{equation}

\noindent In an attempt to further simplify the problem, we assume that the exact structure of the pressure tensor does not play a crucial 
role, and that we can focus on the $x$-component exclusively, $P_{xx}$.
This was verified to be a reasonable assumption by numerical inspection of various conditions.
Also notice that the rotation about $x$ does not modify this component, and $P_{xx} \equiv \tilde{P}_{xx}$.
The problem is further simplified by employing dimensionless variables.
The hydrostatic pressure is computed as $P=\mathrm{tr}(P_{ij})/3$, and non-dimensional variables (superscript $\star$) are obtained as discussed in Eq.~\eqref{eq:dimensionless-variables}, dividing by suitable powers of the density and of the characteristic velocity, 
$\sqrt{P/\rho}$.
Ultimately, our assumptions are such that, for the sake of computing the wave speeds in the $x$ direction, the gas state can be described by only four dimensionless quantities $(P_{xx}^\star,q^\star_x,\tilde{q}_y^\star,\sigma)$.
The dimensionless wave speeds are then approximated as
\begin{subequations}\label{eq:14mom-approx-ws-equations}
\begin{equation}
  \lambda_{\mathrm{max}}^{\star} = \frac{a \sigma + b}{2 \sigma} \left[ \zeta_x + \sqrt{\zeta_x^{2} - 4/5 \zeta_x \sigma C + 4 \sigma^2 Y + \tilde{q}_y^{\star 2}/10} \right] + E  \, ,
\end{equation}
\begin{equation}
  \zeta_x = {q}_x^\star + \sqrt{\tilde{q}_y^{\star 2}} \left( 0.6 \, {P}_{xx}^{\star 2} - 0.38 \, {P}_{xx}^\star + 0.35 \right) \ \  , \ \ \ \tilde{q}_y^\star = (q_y^{\star 2} + q_z^{\star 2})^{1/2} \, , 
\end{equation}
\begin{equation}
  a = 1.4 \, \left({P}_{xx}^\star\right)^{1.1} \ \exp \left[ - {P}_{xx}^{\star 2}\right]  \ , \
  b = 0.9 \, {P}_{xx}^\star \ \exp \left[ - 1/2 \left({P}_{xx}^\star\right)^{1.4}\right] \, ,
\end{equation}
\begin{equation}
   E = \frac{8}{10}\sqrt{(3 - 3\sigma) {P}_{xx}^\star} \ \ \ , \ \ \ \ C = \sqrt{(3 - 3\sigma) {P}_{xx}^\star} \, ,
\end{equation}
\begin{equation}
   B = 5 - 4 \sqrt{\sigma} + \sqrt{10 - 16 \sqrt{\sigma} + 6 \sigma} \ \ \ , \ \ \ 
   Y = B + E^2 - 2 E \sqrt{B} \, .
\end{equation}
\end{subequations}

\noindent The quantity $\zeta_x$ is dimensionless, and does not have a specific physical meaning. 
Instead, it is only meant to embed, empirically, the effect of the transverse heat flux, $q_y$, into the formulation.
The remaining quantities, $a$, $b$, include the non-linear effect of the pressure component, $P^\star_{xx}$, while the 
quantities $B$, $C$, $E$, and $Y$ are slight modifications of the previous definitions of \cite{baradaran2015development}.
The presented fits give the \textit{maximum} dimensionless wave speed in the $x$ direction.
In the frame of the local bulk velocity, this quantity is always positive.
The \textit{minimum} wave speed is obtained from symmetry considerations:
all that one needs to do is to reverse the $x$-component of the heat flux, $q_x^\star$, while all the other moments remain unchanged:
\begin{equation}
  \lambda_{\mathrm{min}}^{\star} = - \lambda_{\mathrm{max}}^{\star}(-q_x^\star) \, .
\end{equation}

\noindent In the frame of the local bulk velocity, $\lambda_{\mathrm{min}}^\star$ is always negative. 
These wave speeds are rough approximations, and it is possible that they under-estimate the real wave speeds in certain regions of the
moment space. 
In case one needs additional stability, a further multiplying parameter, $k \ge 1$, can be employed to inflate them, as $\lambda_\mathrm{max}^\star \to k \lambda_{\mathrm{max}}^\star$.
The final step consists in re-dimensionalizing the wave speeds and 
moving back to the original reference frame, by adding the bulk velocity,
\begin{equation}
  \lambda_{\mathrm{max}} = u_x + \lambda^\star_{\mathrm{max}} \sqrt{P/\rho} \ \ , \ \ \ 
  \lambda_{\mathrm{min}} = u_x + \lambda^\star_{\mathrm{min}} \sqrt{P/\rho} \, .
\end{equation}

\noindent These wave speeds are relative to the $x$-convective flux. 
The wave speeds relative to the $y$ and $z$ directions are obtained by the same considerations.
The simplest numerical strategy consists in implementing this function once, and then calling it separately for the $x$, $y$ and $z$ directions, after swapping the indices.

\begin{figure}[h!tb]
  \centering
  \includegraphics[width=1.0\textwidth]{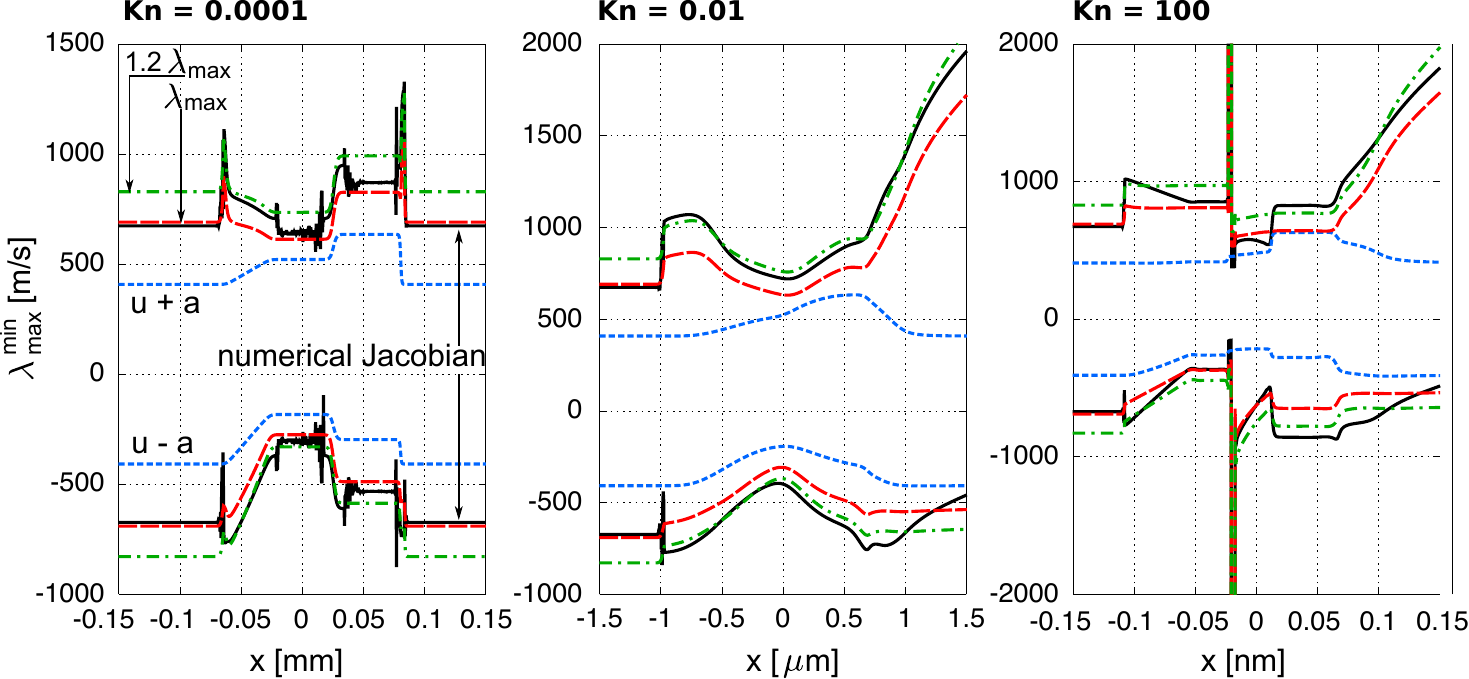}
  \caption{Maximum and minimum wave speeds of the 14-moment system for the Sod shock-tube
problem. Solid black line: numerical wave speeds. Blue dashed line: simple approximation of Eq.~\eqref{eq:wavespeed-u-pm-ka}, with $k=1$.  
        Red dashed line: approximation from Eq.~\eqref{eq:14mom-approx-ws-equations}. Green dot-dashed line:
        further multiplication by a factor of $1.2$.}
  \label{fig:ch-ws-14mom-sod-shock-approx}
\end{figure}

The proposed approximations were obtained primarily by numerical inspection.
In order to check their quality, we have repeated the Sod shock-tube simulations, but this time solving the 14-moment system.
Figure~\ref{fig:ch-ws-14mom-sod-shock-approx} shows that the approximated wave speeds of Eq.~\eqref{eq:14mom-approx-ws-equations} reproduce
very reasonably the numerical wave speeds, and are thus a good improvement over the simpler approximations of Eq.~\eqref{eq:wavespeed-u-pm-ka}.
Some error with respect to the numerical wave speeds is present, and can be mitigated by multiplying the approximated wave speeds by an arbitrary factor, $k$ (in the figure, $k = 1.2$).
As for the 5-moment approximations, employing our approximated wave speeds for highly-rarefied flow problems
might generate some numerical oscillations if certain flux functions are employed, such as the HLL scheme.
The more diffisive Rusanov flux is instead observed to be sufficiently diffusive to damp these oscillations, 
and permits one to employ second-order in space, via linear reconstruction.

\section{2D test case: rarefied jets expanding into a vacuum}\label{sec:two-jets}

In this section, we solve the 14-moment system for two-dimensional problems.
We employ the interpolative maximum-entropy closure, and estimate the wave speeds using the approximated formulas of Section~\ref{sec:ws-14mom-approx-formulas}.
First, in Section~\ref{sec:single-jet}, we study the expansion of a supersonic collisionless (low-density) jet into a vacuum. 
This test case allows us to study the general structure of the solution, and to observe a number of analogies with the 
one-dimensional Sod shock-tube problem (see Section~\ref{sec:sod-shock}).
Then, we consider the problem of two jets, expanding into a vacuum, but crossing each other.
In Section~\ref{sec:two-jets-collisionless}, we study this problem for low-density jets, 
where one expects that no interaction occurs, and the jets re-separate.
Then, in Section~\ref{sec:two-jets-collisional}, we consider a collisional case with a higher density and a finite Knudsen number. 

In all test cases, the domain is a two-dimensional square, with side length $L = 1~\si{m}$.
The jets are centered along the boundaries, have a length $L_\mathrm{jet}=L/7$, and are directed towards the center of the domain.
The jet temperature is fixed at $T_\mathrm{jet} = 300~\si{K}$, and the velocity is chosen to result in a Mach number $\mathrm{M}=5$, or $u_\mathrm{jet}=1613~\si{m/s}$.
All other moments employed in the jet injection are taken at equilibrium.
The jet density is varied among the test cases, and is selected in order to obtain the desired Knudsen number, as discussed in the following.
Numerically, the jets are simulated by imposing the said values into the ghost cells of the domain.
The remaining boundaries of the domain are set to the background density, equal to $\rho_0 = 10^{-5} \rho_{\mathrm{jet}}$, 
that is much lower than the jet density and lets the gas escape.
The same results would be obtained by setting a zero-gradient condition on the fluxes.

Notice that the Mach number is based on the speed of sound, $\mathrm{M} = u/a$.
In the present context, the Mach number is useful, since, at equilibrium, it allows one to uniquely define the problem.
Yet, it should be kept in mind that, especially in collisionless situations, the 14-moment system can predict a set of waves that are faster than the speed of sound.


\subsection{Expansion of a collisionless jet into a vacuum}\label{sec:single-jet}

\begin{figure}[h!tb]
  \centering
  \includegraphics[width=1.0\textwidth]{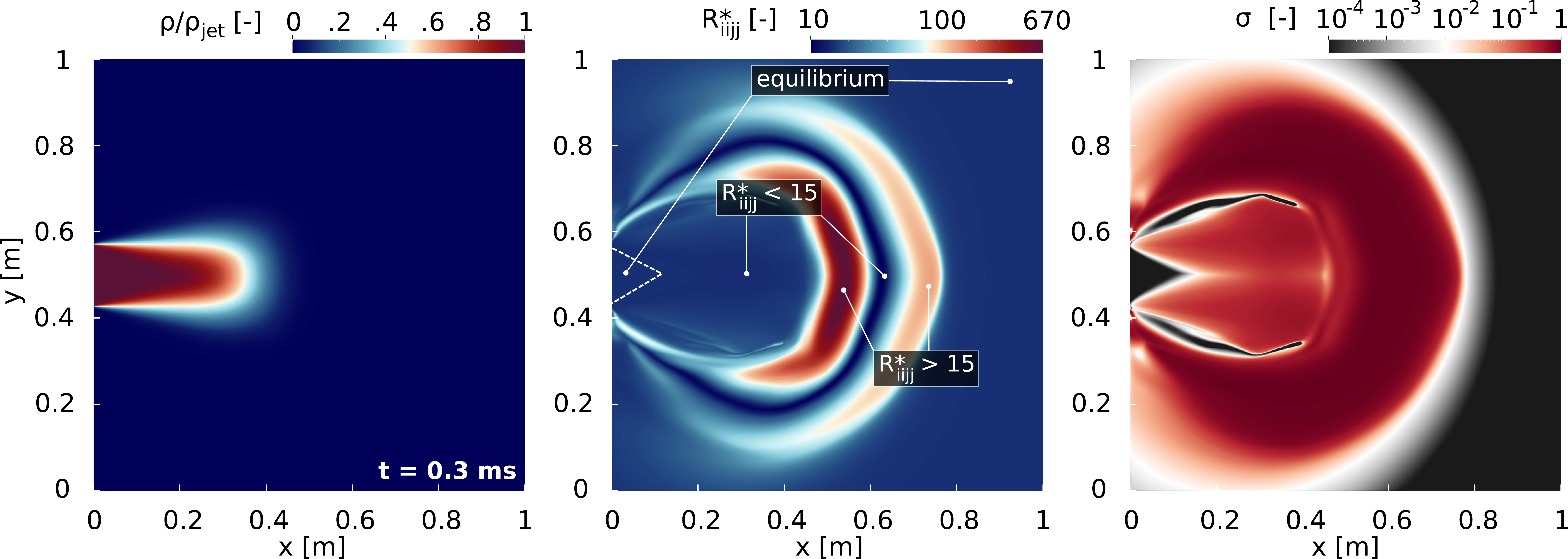}
  \caption{Collisionless jet at a Mach number $\mathrm{M}=5$, expanding into a vacuum. Left: scaled density, linear colormap. Centre: fourth-order dimensionless moment, $R^\star_{iijj}$, logarithmic colormap. Right: parabolic mapping parameter, $\sigma$ ($\sigma=1$ indicates the physical realizability boundary, while $\sigma \to 0$ states are either at equilibrium or near the Junk subspace).}
  \label{fig:single-jet-in-vacuum}
\end{figure}

The solution of the 14-moment maximum-entropy system for a single supersonic and collisionless jet expanding into a vacuum is shown in Fig.~\ref{fig:single-jet-in-vacuum}.
The jet enters the domain from the left, and is shown at a time of $t=0.3~\si{ms}$.
To simulate collisionless conditions, one should employ a sufficiently low density, 
such that the local mean free path is much larger than the domain size.
Here, we directly disable the collision source term instead, and simulate an arbitrary jet density, $\rho_{\mathrm{jet}} = 1$.
Since the solution of the homogeneous system is self-similar with respect to the density, this results in the same 
dimensionless solution. 

The numerical implementation is based on the CUDA version of the Hyper2D solver \cite{boccelli2023hyper2d}, 
and was obtained on an NVIDIA A100 GPU.
The solution is discretized on a grid of $1200\times1200$ elements, with second-order accuracy in space, following the recommendations 
of Section~\ref{sec:MUSCL-for-moments}.
The solution is obtained by marching in time using a first-order forward Euler time-explicit method, imposing a Courant number of $0.5$.
Notice that, since we are neglecting collisions, this solution is self-similar and the same results can be obtained at different time and space scales.

Figure~\ref{fig:single-jet-in-vacuum} shows the scaled density, $\rho/\rho_\mathrm{jet}$, the dimensionless fourth-order moment, $R^\star_{iijj} = R_{iijj}/(P^2/\rho)$,
and the value of the parabolic mapping parameter, $\sigma$.
One can see that the density profile travels roughly at the jet velocity (at the simulated time, $t$, and considering the jet velocity, $u_\mathrm{jet}$, one expects the 
jet to have reached roughly $0.4$--$0.5~\si{m}$).
However, from Fig.~\ref{fig:single-jet-in-vacuum}-Centre, one can see that much faster waves exist.
These waves mostly involve the fourth-order moment, and travel roughly twice as fast, in the present test case.
Lower-order moments are also affected by such fast waves, but to a much lesser extent.
These waves are also discussed in Section~\ref{sec:sod-shock}, for the one-dimensional Sod shock-tube problem.
The background gas that is not yet reached by this fast wave, is still at equilibrium, with $R^\star_{iijj} = 15$ and $q^\star_i = 0$.
As the whole set of waves reaches the equilibrium gas, one observes, phenomenologically, the same pattern observed in the Sod shock-tube problem:
considering Fig.~\ref{fig:single-jet-in-vacuum}-Centre, first the gas is brought to a state of super-Maxwellian kurtosis, where $R^\star_{iijj} > 15$;
then, there is an inversion of this behavior, and a quick decrease in $R^\star_{iijj}$, followed by a further peak, and finally another sub-Maxwellian-kurtosis state,
eventually reaching the unperturbed jet region (potential core of the jet).
In this region, one can verify that $R^\star_{iijj} = 15$ and $q_i^\star = 0$. 
This region is indicated by a white dashed line in Fig.~\ref{fig:single-jet-in-vacuum}-Centre, and is clearly visible in Fig.~\ref{fig:single-jet-in-vacuum}-Right.

The values of $\sigma$ are seen to approach the physical realizability boundary ($\sigma = 1$), indicating the large degree of non-equilibrium of this problem.
From Fig.~\ref{fig:single-jet-in-vacuum}-Right, one can also see two wings where $\sigma$ is very small, and equal to the numerical limiting parameter $\sigma_\mathrm{lim}$,
here set to the value $10^{-4}$.
The gas in these regions is not at equilibrium.
Instead, these regions are located close to the Junk subspace, at $R^\star_{iijj}>15$.
This situation is analogous to what observed in Section~\ref{sec:sod-shock} (consider the points (4) and (5), in Fig.~\ref{fig:sod-5mom-structure-details-rho-q-r-momspace}).


\subsection{Two collisionless crossing jets}\label{sec:two-jets-collisionless}

As discussed, the solution of a \textit{single} supersonic collisionless jet expanding into a vacuum reaches strong degrees of non-equilibrium.
Yet, kinetically, such a problem is rather simple:
in collisionless conditions, the distribution function characterizing this problem is a Maxwellian whose tails are truncated due to geometric effects (finite size of the jet inlet).
As a result, if one was to simulate the same problem using the much simpler Euler equations, one would obtain results that are perhaps inaccurate, 
but nonetheless qualitatively similar.

\begin{figure}[h!tb]
  \centering
  \includegraphics[width=1.0\textwidth]{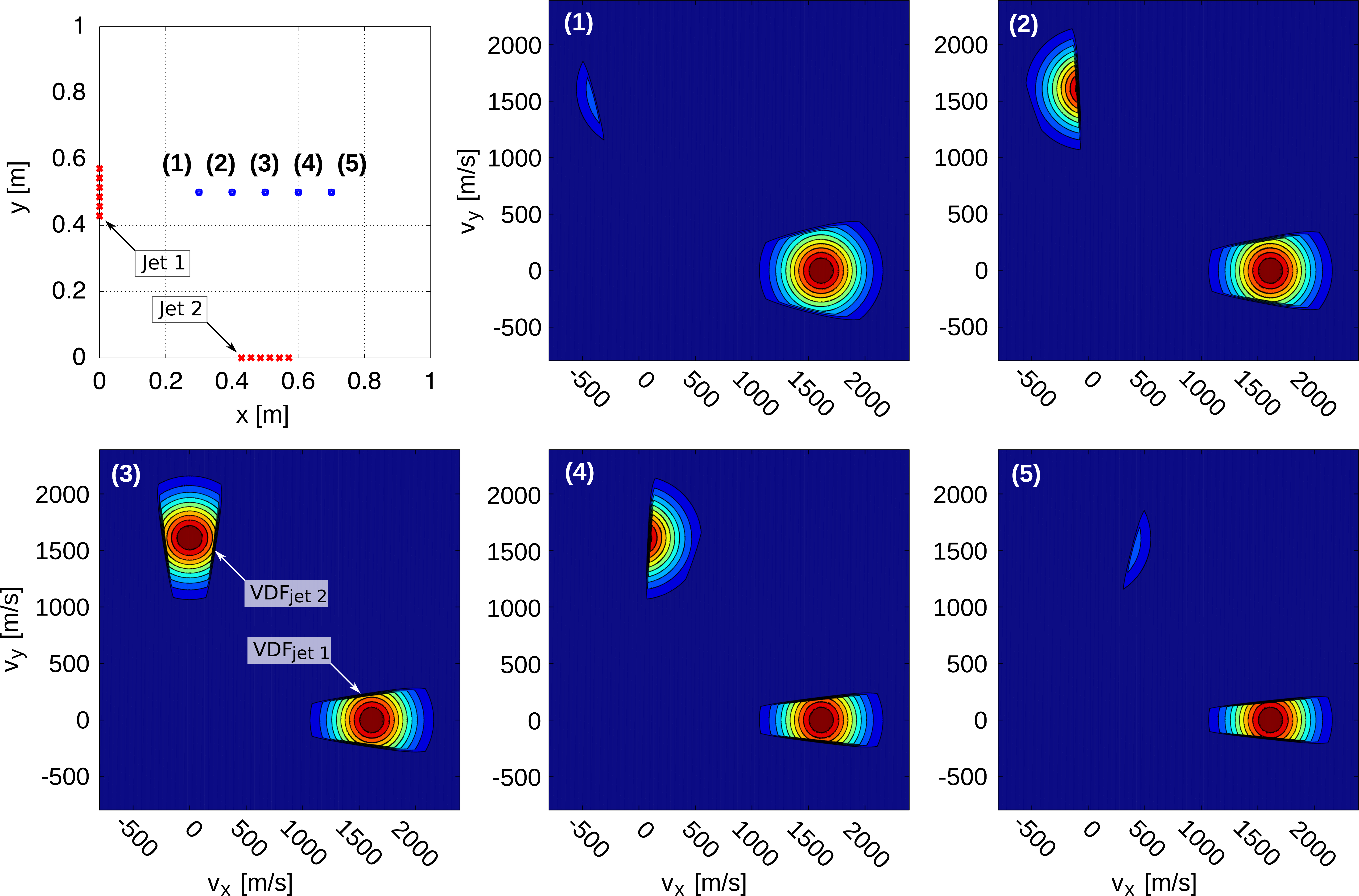}
  \caption{Collisionless crossing jets. Top-Left: computational domain, distribution of point-sources for the kinetic computation (red crosses) and probed points, (1)--(5). 
           Remaining panels: VDF (arbitrary scaling) at these points, in velocity space.}
  \label{fig:ch-maxent-2D-kinetic-jets-VDFs}
\end{figure}

In this section, we complicate the picture. 
We consider two supersonic jets oriented at a right angle, and crossing each other in the middle of the domain. 
This configuration has been frequently employed in the literature, for characterizing moment methods \cite{desjardins2008quadrature,patel2019three}.
Kinetically, this problem is richer than the single-jet case.
This setup, as well as the kinetic solution at various points of the domain, is described in Fig.~\ref{fig:ch-maxent-2D-kinetic-jets-VDFs}.
The VDFs shown in this figure are a semi-analytical solution of the collisionless kinetic equation, and are obtained by first discretizing the jets into a number of point sources, 
and then adding each individual contributions, accounting for the view factor.
The reader is referred to \cite{boccelli2021moment} for a detailed discussion of the numerical solution, and to \cite{kogan1969rarefied} for analogous analytical results.
In the collisionless regime, the kinetic solution is simply the superposition of the two independent beams of particles.

Notice that, while physically simple, this problem cannot be reproduced using traditional fluid models.
Indeed, in order to predict the re-separation of the jets, it is necessary that the system of equations can reproduce states associated with bi-modal VDFs (see Fig.~\ref{fig:ch-maxent-2D-kinetic-jets-VDFs}).
Traditional gas dynamic models, such as the Euler equations, would instead always predict an interaction of the jets, 
regardless of the collisionality \cite{forgues2019higher}. 
The solution of the Euler equations for this problem is discussed in \ref{sec:appendix-euler-jets}.

\begin{figure}[h!tb]
  \centering
  \includegraphics[width=1.0\textwidth]{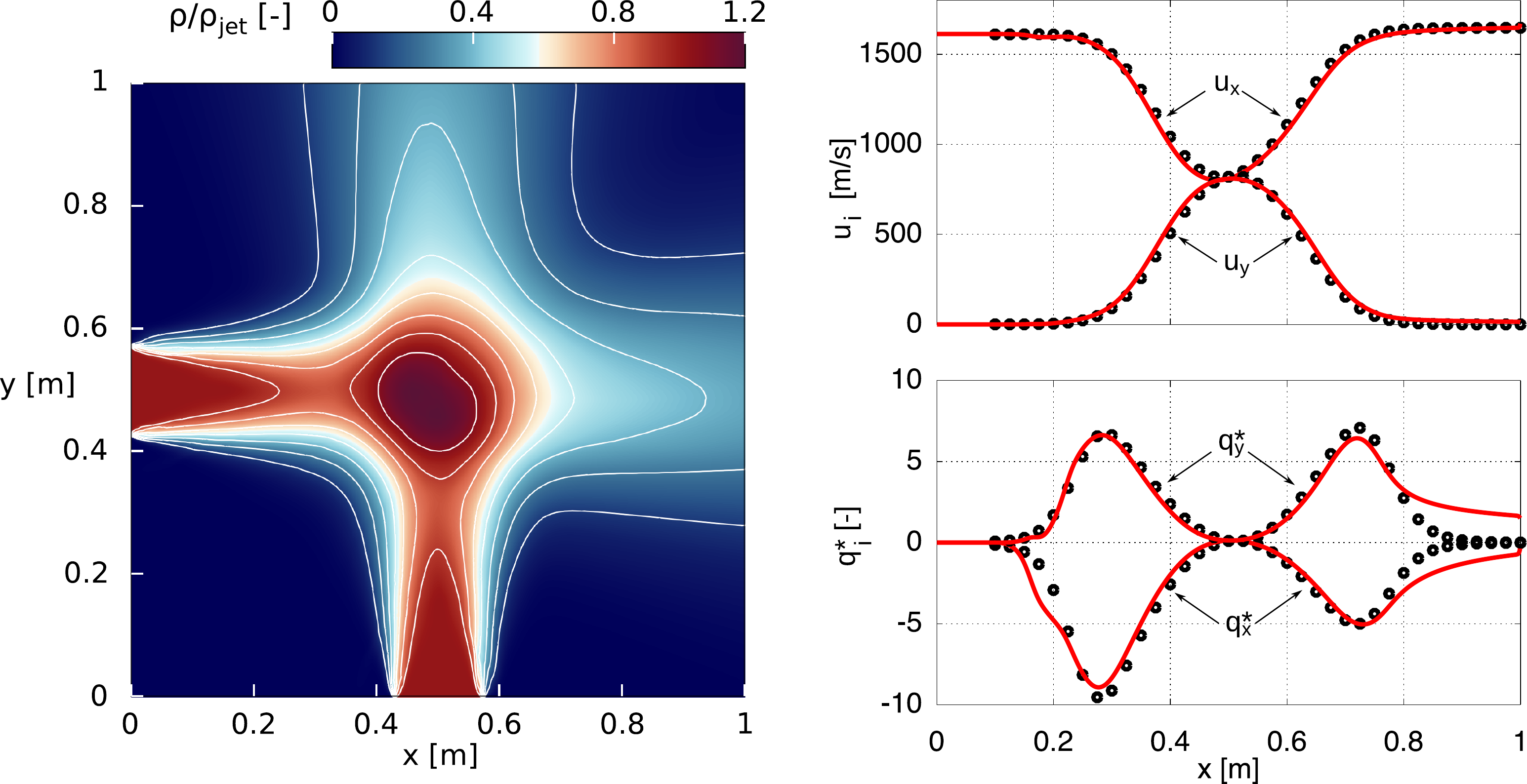}
  \caption{Collisionless crossing jets. Left: solution of the 14-moment system (density). 
           Right: comparison with selected moments, obtained from the kinetic VDFs, along the axis of the left-jet (line $y = 0.5$). 
           Solid line: maximum-entropy solution. Symbols: kinetic solution.}
  \label{fig:ch-maxent-2D-14mom-jets-density}
\end{figure}

\begin{figure}[h!tb]
  \centering
  \includegraphics[width=1.0\textwidth]{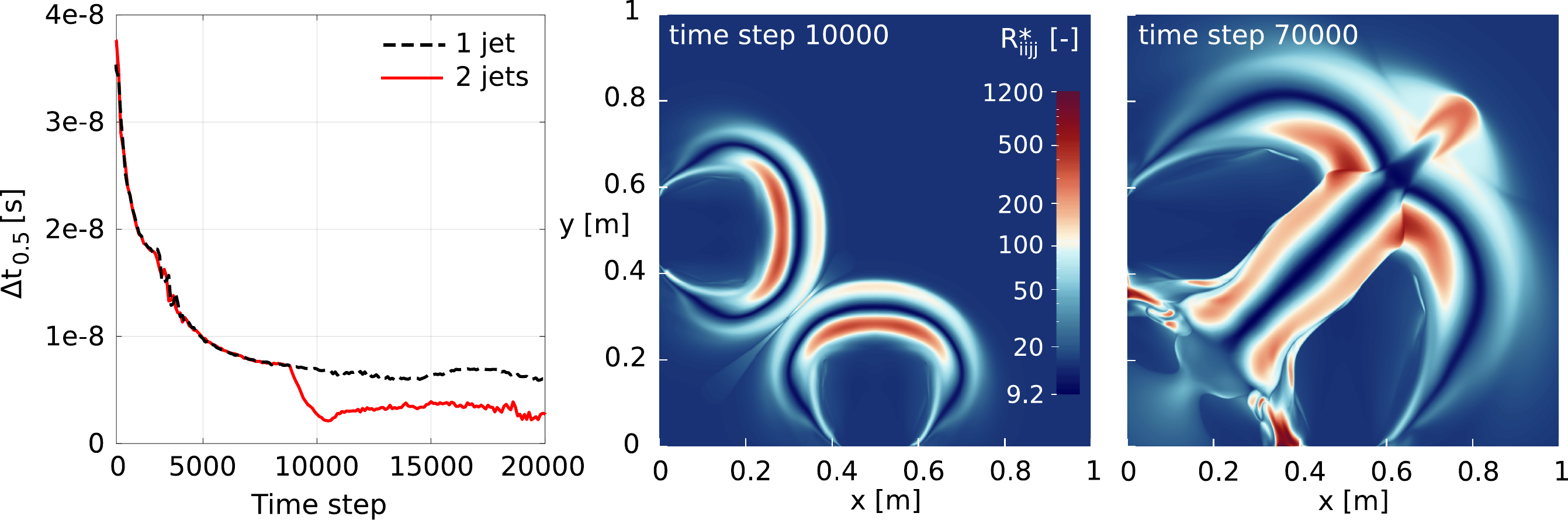}
  \caption{Collisionless crossing jets. Left: comparison of the maximum timestep $\Delta t$ allowed by the CFL constraint, in the single-jet and in the two-jet simulations. Centre and Right: dimensionless moment $R^\star_{iijj}$ at two different time steps.}
  \label{fig:ch-maxent-2D-1jet-2jets-comp-cost}
\end{figure}

Figure~\ref{fig:ch-maxent-2D-14mom-jets-density}-Left shows the solution of the 14-moment maximum-entropy system, in absence of collisions.
The grid and numerical methods are the same as for the test case of Section~\ref{sec:single-jet}, and the solution is obtained by marching in time until convergence.
The 14-moment method is successful in reproducing the jet re-separation.
In Fig.~\ref{fig:ch-maxent-2D-14mom-jets-density}-Right, we show some selected moments, extracted along the centerline of the horizontal jet, 
and we compare them to the moments obtained from the semi-analytical VDFs. 
Remarkably, the maximum-entropy solution shows a good accuracy in most of the domain, despite the complete lack of collisions and the consequently high non-equilibrium.
In terms of computational cost, one observes that the crossing-jet test case is more demanding than the previous single-jet case.
As expected, as long as the jets are independent, the two computational costs are the same (see Fig.~\ref{fig:ch-maxent-2D-1jet-2jets-comp-cost}).
However, when the fast waves of the first and the second jet interact (approximately at time step 9000, for our grid), 
the two-jet solution climbs the Junk subspace, and a smaller time step is necessary in order to satisfy the CFL condition.
It should be stressed that, whereas the fast waves interact and create a rich shock structure (see Fig.~\ref{fig:ch-maxent-2D-1jet-2jets-comp-cost}-Centre and Right),
this has little effect on the density field, and the bulk of the jets simply re-separates.


\subsection{Collisional crossing jets}\label{sec:two-jets-collisional}

\begin{figure}[h!tb]
  \centering
  \includegraphics[width=1.0\textwidth]{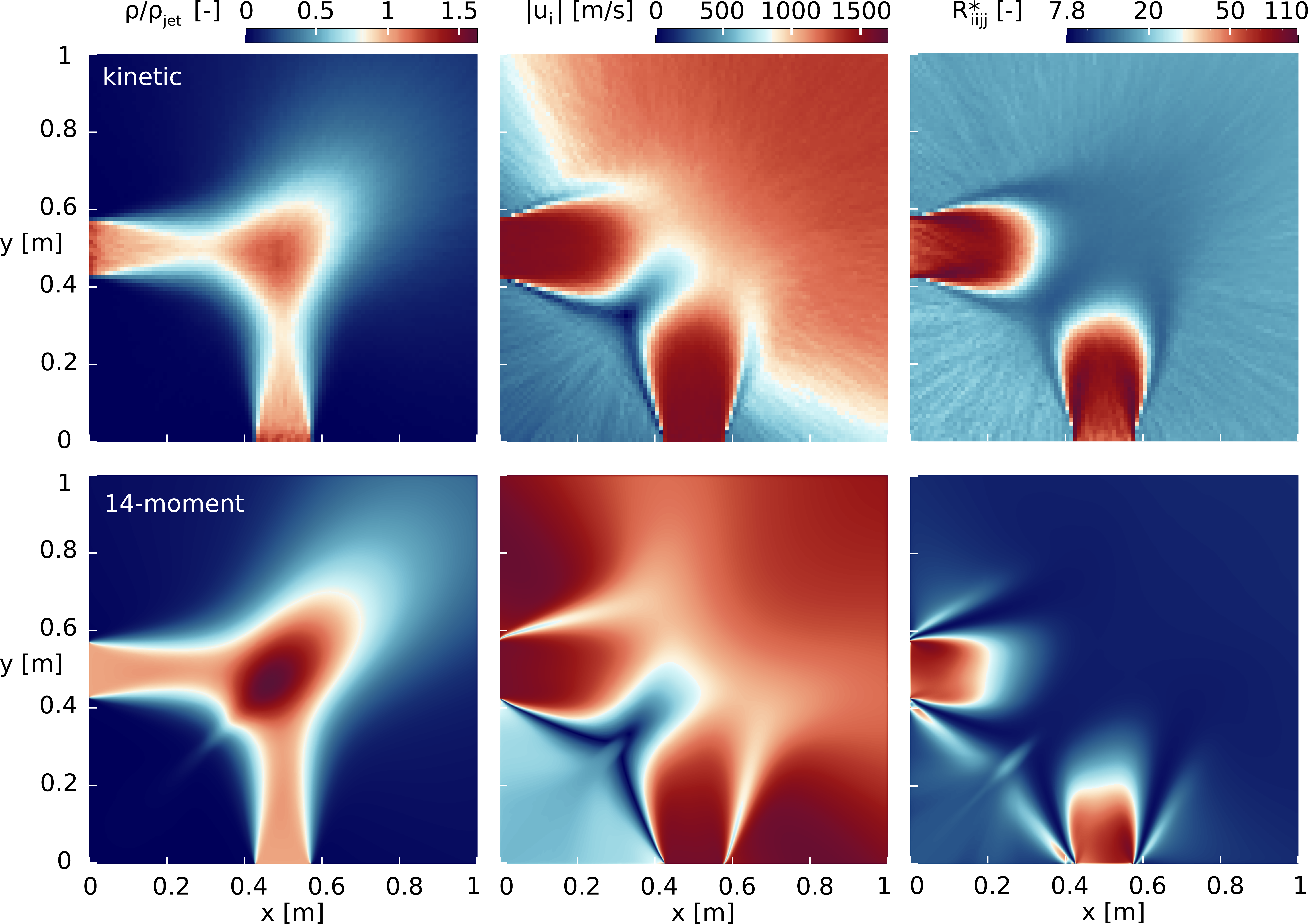}
  \caption{Comparison of a particle-based kinetic solution (Top panels) and a solution of the 14-moment system (Bottom) for the rarefied crossing-jet
           problem at $\mathrm{Kn}=1$. The panels show the scaled density, $\rho/\rho_\mathrm{jet}$, the absolute value of the velocity, $|u_i|$, and the dimensionless
           fourth-order moment, $R^\star_{iijj}$. All these fields are shown at steady state.}
  \label{fig:ch-maxent-2D-vs-kin-collisions}
\end{figure}

As the jet density is increased, the two streams of particles begin to interact.
Figure~\ref{fig:ch-maxent-2D-vs-kin-collisions} shows the result of a collisional simulation, obtained at $\mathrm{Kn}=\lambda_\mathrm{jet}/L_\mathrm{jet}=1$,
where $L_\mathrm{jet}$ is the jet size and $\lambda_\mathrm{jet} = (\sqrt{2}\, n_\mathrm{jet}\, \sigma)^{-1}$ is the mean free path in the unperturbed jet, with
$n$ the number density and where $\sigma=5.463\times10^{-19}~\si{m^2}$ is a constant elastic cross-section.
At $\mathrm{Kn}=1$, the jet density is $\rho_\mathrm{jet} = 6.01\times10^{-7}~\si{kg/m^3}$.
The collision frequency appearing in the BGK collision operator, $\nu$, should depend on the shape of the distribution function, and thus on 
non-equilibrium.
However, as done in the study of the shock-tube problem, we consider here a simplified expression that does not depend on the 
particle velocity, but only on the local macroscopic fields,
$\nu = v^{\mathrm{th}}/\lambda$, where $v^{\mathrm{th}} = \sqrt{8 k_B T/\pi m}$ is the thermal velocity for a Maxwellian distribution.

A kinetic solution was also obtained, and is compared to the 14-moment solution in Fig.~\ref{fig:ch-maxent-2D-vs-kin-collisions}.
Due to the presence of collisions, this problem does not have a semi-analytical solution.
Instead, we employ here a particle method, implemented in an in-house code (see \cite{boccelli2023modeling}).
The solution follows the DSMC (Direct Simulation Monte Carlo) method \cite{bird1994molecular} for what concerns particle injection and streaming, 
while collisions are performed using the BGK operator, as described in the following.
First, the average density, velocity and temperature are computed on the grid cells. 
Every particle is then checked for collisions, employing a collision frequency, $\nu$, calculated from the density and temperature in the local cell.
For every particle, a uniformly distributed random number, $\mathfrak{R}\in[0,1]$, is extracted, and the collision is performed if $\mathfrak{R} \le P_\mathrm{coll}$,
with
\begin{equation}
  P_\mathrm{coll} = 1 - \exp(- \nu \Delta t) \, .
\end{equation}

\noindent If the collision happens, the particle velocity is re-sampled from a Maxwellian VDF, at the local density, bulk velocity and temperature.
Effectively, in the BGK model, all particles collide with the same probability, with a background of equilibrium particles.
More accurate collision models are possible, including the DSMC method itself, or the adoption of velocity-dependent collision frequencies \cite{cercignani1988boltzmann}.
However, we select the BGK model to ensure consistency with the current 14-moment formulation. 
For other examples of particle-based solutions of the BGK collision operator, the reader is referred to \cite{gallis2000application,macrossan2001particle}.

At $\mathrm{Kn}=1$, the mean free path in the jet core, $\lambda_\mathrm{jet}$, is equal to the jet size.
We select a grid, for the particle simulation, composed of $100\times100$ cells.
This results in a cell size $\Delta x \approx \lambda_\mathrm{jet}/14$.
From an analysis of the simulation results, it can be confirmed that this grid resolves the mean free path everywhere in the domain.
The time step is selected such that the maximum collision probability, among all particles, does not exceed the value $P_{\mathrm{coll}}^\mathrm{max} = 0.2$.
The number of simulated particles is approximately 1 million.
As a result, every cell in the path of the jets contains more than $10$ particles. 
Some regions in the domain do not respect this requirement (consider for instance the bottom-left corner, between the two jets).
In order to fill these cells, one should employ a much larger number of particles, but with little or no effect on the overall simulation, since these regions are associated with an extremely low density anyway.
The particle solution shown in Fig.~\ref{fig:ch-maxent-2D-vs-kin-collisions} is obtained by averaging the fields in time, after reaching convergence. 
Approximately $2000$ time steps are employed for the post processing average, each time step being spaced by $5$ time steps from the previous one, to reduce 
possible correlations.
Note that time-ensemble averages are particularly important for higher-order moments, that are strongly affected by statistical noise.

It should be remarked that the collision method employed here conserves the mass exactly,
but only conserves the momentum and energy statistically.
In closed-system simulations, this implies that the temperature may artificially change in time. 
In the present case, we employ an open configuration and this issue turns out to be negligible:
the jets constantly introduce new particles into the domain, and the numerical error remains bounded in time.

With respect to the collisionless test case, the Kn=1 case shows some initial interaction of the jets.
From the density plot, one can see the two jets merging into a single diffused profile.
The diffusion appears to be partially under-reproduced by the maximum-entropy method, 
and the peak density is over-predicted.

At a first glance, the maximum-entropy velocity field might appear to differ significantly from the particle solution.
However, it should be considered that most of the deviation occurs out of the jet cores and away from 
the interaction region.
In this part of the domain, the physical solution is characterized by a very low density.
This has the effect of increasing the velocity, in the maximum-entropy simulation.
In the rest of the domain, where the density is larger, or where the jets interact, 
the maximum-entropy solution shows a reasonable agreement with the kinetic solution.
The maximum-entropy method manages to reproduce the main velocity-field features, 
such as the shape of the jet lobes and the low-velocity region that borders the jets.
The fourth-order moment plot shows the largest discrepancies from the kinetic solution.
As shown, this field is characterized by a number of discontinuities, which are not present in the physical solution.




\section{Conclusions}

In this work, we have investigated the application of the 5 and 14-moment maximum-entropy systems to gases in rarefied and supersonic conditions.
The Sod shock-tube problem has been studied at various degrees of rarefaction, and the kinetic solution has been compared to a solution of the maximum-entropy system.
In rarefied conditions, the solution of the shock-tube problem deviates from thermodynamic equilibrium, and spans the whole moment space. 
The maximum-entropy systems proved to be able to reproduce this solution up to large values of the Knudsen number, although with a varying accuracy. 
In the fully collisionless regime, the maximum-entropy system predicts a set of discontinuities, a smooth shock profile and a rarefaction fan.
The trajectory of such features in moment space is discussed.
It is observed that some of the discontinuities bring the system very close to the Junk singularity, with a consequent increase in the computational time.

A two-dimensional problem was also studied, consisting in crossing supersonic jets at a Mach number $\mathrm{M}=5$, and expanding into a vacuum.
The kinetic solution was compared to the solution of the 14-moment system, that was shown to be remarkably accurate.
The numerical simulation of the maximum-entropy system at high degrees of rarefaction and in supersonic conditions requires 
numerical care. 
In this work, we discuss the choice of the numerical interface fluxes, and indicate how to obtain a robust second-order spacial discretization.
Also, we provide approximated expressions for the maximum and minimum wave speeds of the 14-moment system.
These approximations allow one to avoid the costly numerical computation of the eigenvalues of the flux Jacobian for the system of 14 equations, and permit to
accelerate the simulations considerably.
These approximations are relevant to a number of collisionless situations, including the simulation of (low-density) plasmas \cite{boccelli202214}.

Two-dimensional simulations are accelerated on Graphics Processing Units (GPUs), with CUDA.
Our simple CUDA implementation of the maximum-entropy system, not implementing any advanced memory management strategy, resulted in a speed-up of 
roughly 340 times, when run on an NVIDIA A100 GPU, with respect to single-core computations.
Given the complex structure of the maximum-entropy closure, involving a considerable number of variables, older GPUs tend to fill all available registers.
This requires either a careful implementation, with register re-using, or to adopt a sufficiently low number of CUDA threads per block, in the kernel launch configuration 
($8\times8$ threads per block showed to work fine on both the A100 and on a Tesla K20X GPU).


\section*{Acknowledgements}

SB would like to thank Prof. Aldo Frezzotti (Politecnico di Milano) for the enlightening discussions on the topic of this work.
Funding for this project was provided by the Natural Sciences and Engineering Research Council of Canada (NSERC)
through grant number RGPAS-202000122. 
This research was supported by grants from NVIDIA and utilized NVIDIA A100 40GB GPUs.
The authors are very grateful to NSERC and NVIDIA for this support.

\appendix

\section{Artificial limiting of $\sigma$ near the Junk line}\label{sec:appendix-sigma-lim}

As discussed in Section~\ref{sec:max-ent}, the fourth-order maximum-entropy method is affected by the presence of a singularity region, 
where the closing flux diverge, and so do the wave speeds.
This presents numerical challenges.
For one, the Courant number associated with time-esplicit computations would become impractically small.
In this work, during the numerical calculations, we adopt the following procedure:
the closing fluxes are computed from Eq.~\ref{eq:closure-1D1V-s-approx} (for the one-dimensional case) and 
Eqs.~\ref{eq:closure-14mom-Qijk-approx}, \ref{eq:closure-14mom-Rijkk-approx} and \ref{eq:closure-14mom-Sijjkk-approx} (three-dimensional case), 
employing a value $\bar{\sigma} = \mathrm{max}(\sigma, \sigma_\mathrm{lim})$, where $\sigma_\mathrm{lim}$ is a user-defined limitation parameter, 
typically chosen in the range of $10^{-4}$ to $10^{-5}$.
The other moments appearing in the closure equations are unchanged.

\begin{figure}[h!tb]
  \centering
  \includegraphics[width=0.4\textwidth]{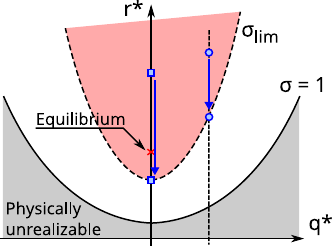}
  \caption{Effect of the limiting of $\sigma$. The limiting region, $\sigma < \sigma_\mathrm{lim}$ is shown in red, and is exaggerated to aid the representation.
           The blue symbols and arrow represent the limitation acting on two hypothetical gas states.}
  \label{fig:limiting-procedure}
\end{figure}

Figure~\ref{fig:limiting-procedure} shows the implications of this limiting procedure, for the one-dimensional case:
if the gas state is inside the red-shaded region limiting is activated and the closing flux is computed from a state that has $\sigma = \sigma_\mathrm{lim}$ 
and the same heat flux, $q^\star$.
With this strategy, one is artificially reducing the value of the fourth-order moment, $r^\star$.
It should be stressed that the limitation is only performed, internally, in the subroutine that computes the closing 
fluxes and the wave speed:
no permanent modification is imposed on the actual state of the gas inside the solution vector.
As a result, the solution vector can approach the Junk line freely, but the closing fluxes are computed 
from a (temporary) $\sigma$-limited state.

Other approaches are possibile. 
For instance, one could consider pushing the state to the $\sigma_\mathrm{lim}$ parabola horizontally, rather than vertically.
This would corresponds to keeping $r^\star$ constant, and modifying $q^\star$. 
However, this strategy would not be well defined for symmetric states located on the line $q^\star = 0$.
Ultimately, the strategy proposed in our work appears to be robust for all considered test cases.

\section{Development of the approximated wave speeds}\label{appendix:approx-wave-speeds}

In this section, we formulate an approximation for the maximum and minimum wave speeds of the 14-moment maximum-entropy system.
As a starting point, we consider the much simpler 1D1V 5-moment maximum-entropy system.
Notice that, in the limit of a vanishing temperature in two out of the three spacial directions, the 14-moment system collapses to the 1D1V 5-moment formulation.
Therefore, it makes sense to take this as a starting point.

\begin{figure}[h!tb]
  \centering
  \includegraphics[width=0.5\textwidth]{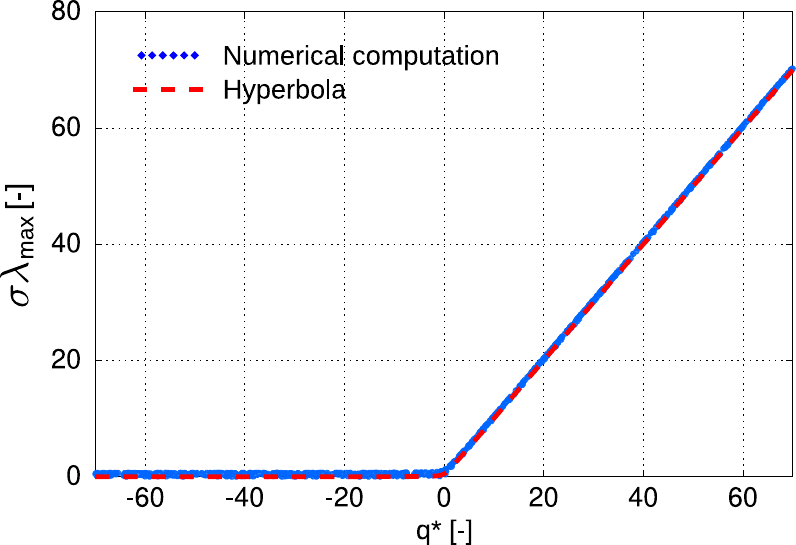}
  \caption{Maximum wave speed for the 5-moment system. The blue dots represent numerical eigenvalue computations from the 5-moment Jacobian. The red dashed line is the 
           simple rectangular hyperbola approximation of Eq.~\ref{eq:ch-maxent-5mom-hyperbola-approx-simplest}. 
           By plotting the quantity $\sigma \lambda_\mathrm{max}^\star$, instead of $\lambda_\mathrm{max}^\star$ alone, all points roughly collapse on the hyperbola.}
  \label{fig:ch-maxent-5mom-ws-q-hyperbola}
\end{figure}

As shown in Fig.~\ref{fig:ch-maxent-5mom-ws-q-hyperbola}, the maximum wave speed of the 5-moment system, when multiplied by $\sigma$ and plotted against $q^\star$, roughly follows a rectangular hyperbola,
\begin{equation}\label{eq:ch-maxent-5mom-hyperbola-approx-simplest}
  \sigma \lambda_{\mathrm{max}}^{\star (5)}  \approx \frac{1}{2} \left( q^\star + \sqrt{q^{\star 2} + 1}\right) \, .
\end{equation}

Dealing with the quantity $\sigma \lambda$ (rather than $\lambda$ alone) is convenient, 
as this removes the singularity in the wave speeds that diverge as the Junk line is approached ($\sigma \to 0$).
The hyperbola approximation reproduces the general behavior.
However, there are two main inaccuracies:
\begin{itemize}
  \item The values near equilibrium are not exact, and should include a dependence on $\sigma$;
  \item For negative values of $q^\star$, the real wave speed is not zero, and a vertical shift is necessary.
\end{itemize}

\noindent More refined formulas have been developed by Baradaran \cite{baradaran2015development}:
\begin{equation}\label{eq:ws-max-baradaran-orig}
  \lambda_{\mathrm{max}}^{\star (5)} = \frac{1}{2 \sigma}\left[ q^\star + \sqrt{q^{\star 2} - \frac{4}{5} q^\star \sigma C + 4 \sigma^2 Y }\right] + E \, ,
\end{equation}

\noindent with
\begin{subequations}
  \begin{equation}
  E = 8/10\sqrt{3 - 3 \sigma}  \ \ , \ \ B=5-4\sqrt{\sigma}+\sqrt{10-16\sqrt{\sigma}+6\sigma} \, ,
  \end{equation}
  \begin{equation}
  C=\sqrt{3 - 3\sigma} \ \ , \ \ Y=B+E^2-2E\sqrt{B} \, .
  \end{equation}
\end{subequations}

\noindent This approximation is not exact, but includes further dependence on $\sigma$ in the hyperbolic part, and a vertical shift (through the factor $E$, that also depends on $\sigma$).
We start from Baradaran's approximation for building the 14-moment wave speed estimates.


\subsection{Axisymmetric 14-moment system}

Before extending the approximation to the full 14-moment case, it is useful to consider an intermediate case.
To simplify our analysis, we consider first a spacially axisymmetric case, where quantities change only along $x$.
In \cite{mcdonald2013affordable}, McDonald \& Torrilhon discuss the full axisymmetric maximum-entropy system, that is composed of six independent equations.
In the directions perpendicular to $x$, one has $u_y = u_z = 0$ and $q_y = q_z = 0$.
The axisymmetric pressure tensor is written as 
\begin{equation}
  P_{ij} = 
  \begin{bmatrix}
    P_{xx} & 0 & 0 \\
    0 & P_{rr}/2 & 0  \\
    0 & 0 & P_{rr}/2  \\
  \end{bmatrix} \, ,
\end{equation}

\noindent where $P_{rr} = P_{yy} + P_{zz}$, the index $r$ indicating the radial direction.
In this model, the gas state is fully defined by the following six quantities: $\rho$, $u_x$, $P_{xx}$, $P_{rr}$, $q_x$ and $R_{iijj}$.
Non-dimensionalization is performed following Eq.~\eqref{eq:dimensionless-variables}, 
and $P_{ij}^\star = P_{ij}/P$, where $P = (P_{xx} + P_{rr})/3$.
At equilibrium, one has $P_{xx}^\star = 1$ and $P_{rr}^\star = 3 - P_{xx}^\star = 2$.
After non-dimensionalizing and moving to the frame of the bulk velocity, the thermodynamic state is fully defined by three variables:
$P_{xx}^\star$, $q_x^\star$ and the mapping parameter, $\sigma$ (or alternatively $R_{iijj}$).
Therefore, with respect to the 1D1V case, the three-dimensional axi-symmetric case (1D3V) only introduces one additional variable, $P_{xx}^\star$.

\begin{figure}[h!tb]
  \centering
  \includegraphics[width=1.0\textwidth]{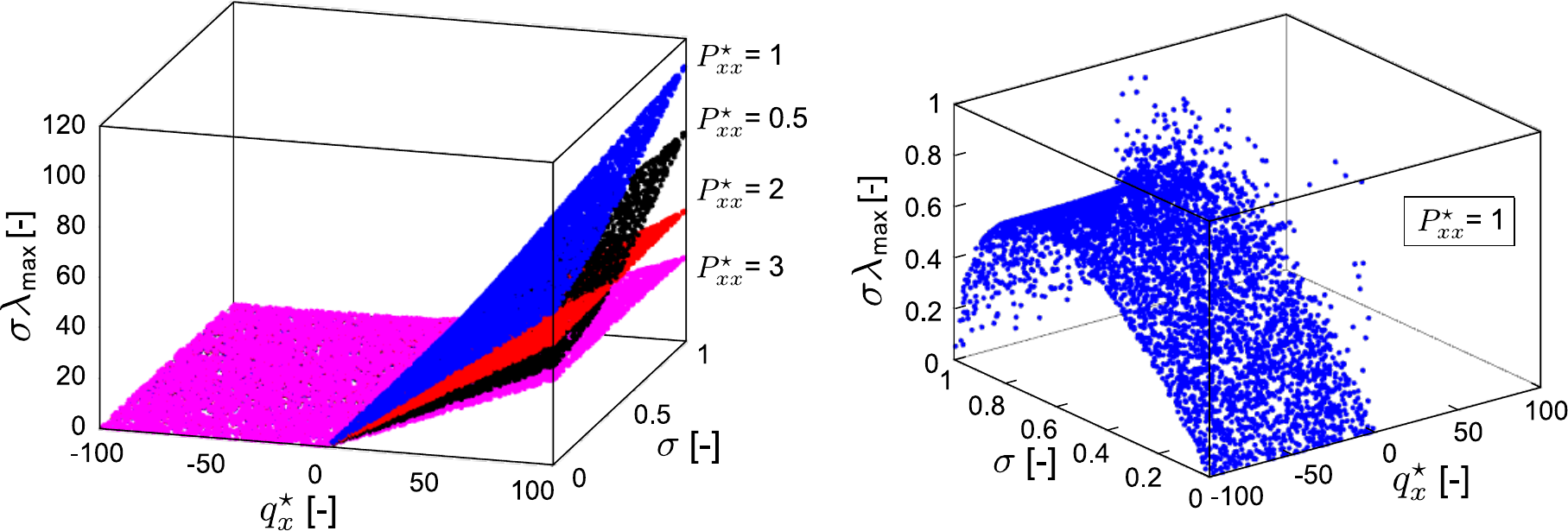}
  \caption{Maximum wave speed, scaled by $\sigma$, for the axisymmetric 14-moment system. 
           Left: wave speed sampled at random points in the $q_x^\star-\sigma$ plane, 
           for prescribed values of $P_{xx}^\star$. 
           Right: magnification around the region of small $q_x^\star$, with $P_{xx}^\star = 1$.} 
  \label{fig:ch-maxent-ws-6mom-randpoints}
\end{figure}

In order to assess the influence of $P_{xx}^\star$ on the wave speeds of the axisymmetric 14-moment system, 
we perform a set of numerical eigenvalue calculations.
Figure~\ref{fig:ch-maxent-ws-6mom-randpoints} shows a numerical inspection of the maximum wave speed for this system, 
in the $q_{x}^\star-\sigma$ plane and for different values of $P_{xx}^\star$.

We observe that varying $P_{xx}^\star$ has an effect on the slope of the hyperbola, and (although not shown in
the image) there is an effect in the vertical shift as well.
Therefore, we postulate that the modified wave speeds take the form 
\begin{equation}\label{eq:ws-max-baradaran-modified}
\sigma \lambda_{max} \approx W(P_{xx}^\star, \sigma)\frac{1}{2}\left[ q_x^\star 
                               + \sqrt{q_x^{\star 2} - \frac{4}{5} q^\star \sigma C + 4 \sigma^2 Y }\right] 
                   + \chi(P_{xx}^\star) \sigma E \, ,
\end{equation} 

\noindent where two additional functions have been introduced, $W(P_{xx}^\star,\sigma)$ and $\chi(P_{xx}^\star)$.
The latter function can be found by considering the limit of $q_x^\star \to -\infty$, 
where Eq.~\ref{eq:ws-max-baradaran-modified} becomes
\begin{equation}\label{eq:lim-6mom-qminusinf}
\lim_{ q_x^\star\rightarrow-\infty } 
\sigma \lambda_{\mathrm{max}} = \chi(P_{xx}^\star) \, \sigma \, E \, .
\end{equation}

\noindent By numerical inspection, it is found that the choice $\chi(P_{xx}^\star) = \sqrt{P_{xx}^\star}$ 
fits well the results, in the whole range, $P_{xx}^\star \in [0, 3]$. 

\begin{figure}[h!tb]
  \centering
  \includegraphics[width=0.5\textwidth]{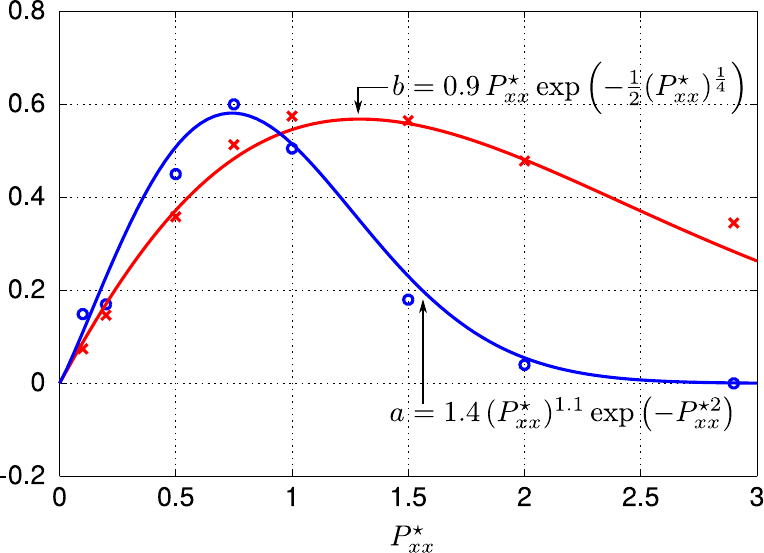}
  \caption{Fit of the slope $a$ and intercept $b$ as a function of the pressure. 
           Symbols: values obtained from the numerical eigenvalues. Solid lines: our numerical fits.}
  \label{fig:ch-maxent-ws-6mom-slopes-a-b}
\end{figure}

We then move to the limit of large $q_x^\star$.
As shown in Fig.~\ref{fig:ch-maxent-ws-6mom-randpoints}, the wave speed settles to planes whose slope and intercept
depend on $P_{xx}^\star$.
in other words, for a constant value of $q_{x}^\star$, the quantity $\sigma \lambda_\mathrm{max}$ is a straight line,
if plotted against $\sigma$.
Therefore, we write
\begin{equation}
  W(P_{xx}^\star) = a(P_{xx}^\star) \, \sigma + b(P_{xx}^\star) \, ,
\end{equation}

\noindent and we identify some suitable forms for $a(P_{xx}^\star)$ and $b(P_{xx}^\star)$ by numerical inspection,
as shown in Fig.~\ref{fig:ch-maxent-ws-6mom-slopes-a-b}.

\begin{figure}[h!tb]
  \centering
  \includegraphics[width=1.0\textwidth]{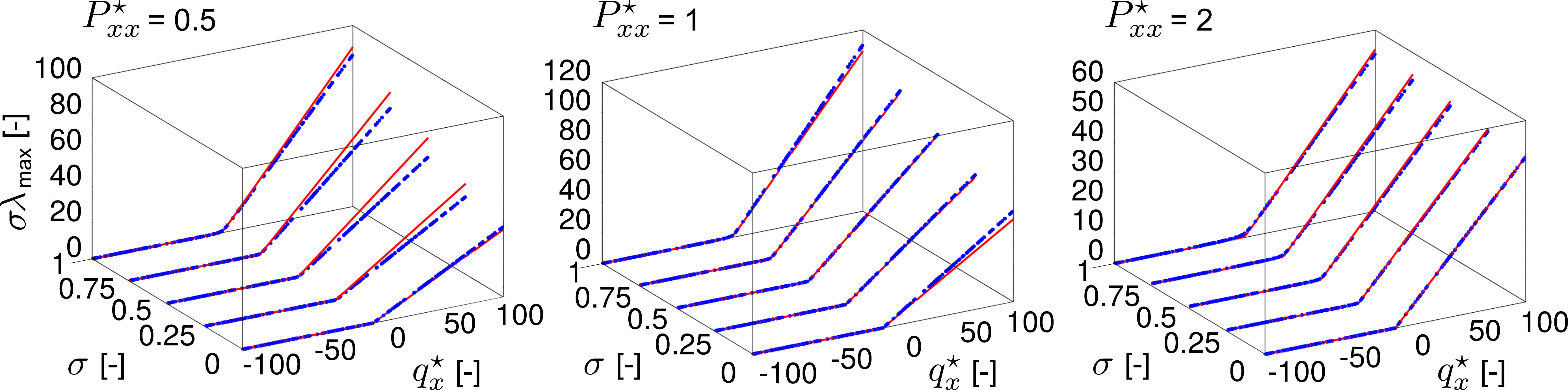}
  \caption{Maximum wave speed for the 6-moment system. Red lines: approximated values from Eq.~\eqref{eq:6mom-approx-ws-equations}. 
           Blue dots: numerically computed values.}
  \label{fig:ch-maxent-ws-6mom-check-wholescale}
\end{figure}

\begin{figure}[h!tb]
  \centering
  \includegraphics[width=1.0\textwidth]{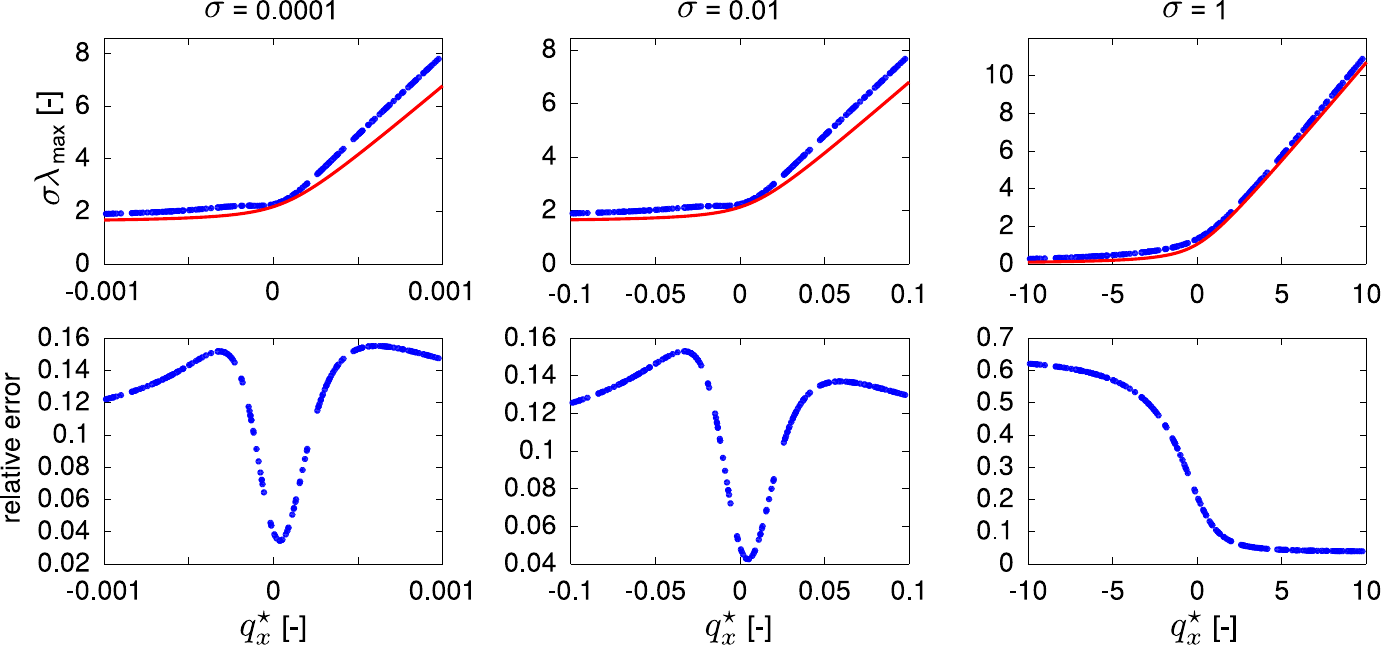}
  \caption{Maximum wave speed for the 6-moment system (Top) and relative error (Bottom) for $P_{xx}^\star=1$.
           Red lines: approximated values from Eq.~\eqref{eq:6mom-approx-ws-equations}.
           Blue dots: numerically computed values.
           Bottom: relative error.}
  \label{fig:ch-maxent-ws-6mom-check-zoom}
\end{figure}

Ultimately, the maximum wave speed is approximated as

\begin{subequations}\label{eq:6mom-approx-ws-equations}
\begin{equation}
  \lambda_{\mathrm{max}}^{(6)\star} = \frac{a \sigma + b}{2 \sigma} \left[ q_x^\star + \sqrt{q_x^{\star 2} - 4/5 q_x^\star \sigma C + 4 \sigma^2 Y} \right] + E  \, ,
\end{equation}
\begin{equation}
  a = 1.4 \, \left(P_{xx}^\star\right)^{1.1} \ \exp \left[ - P_{xx}^{\star 2}\right]  \ , \ 
  b = 0.9 \, P_{xx}^\star \ \exp \left[ - 1/2 \left(P_{xx}^\star\right)^{1.4}\right] \, ,
\end{equation}
\begin{equation}
   E = \frac{8}{10}\sqrt{(3 - 3\sigma) P_{xx}^\star} \ \ \ , \ \ \ \ C = \sqrt{(3 - 3\sigma) P_{xx}^\star} \, ,
\end{equation}
\begin{equation}
   B = 5 - 4 \sqrt{\sigma} + \sqrt{10 - 16 \sqrt{\sigma} + 6 \sigma} \ \ \ , \ \ \ 
   Y = B + E^2 - 2 E \sqrt{B} \, .
\end{equation} 
\end{subequations}

\noindent where some additional minor modifications are introduced---such as a factor $\sqrt{P_{xx}^\star}$ in $C$---in attempt to improve the accuracy.
Clearly, this formulation is emipirical and heavily approximated, as it adds our inaccuracy on top of that 
already present in Baradaran's approximation.
Better formulas should be employed, if available.
Yet, as shown in Fig.~\ref{fig:ch-maxent-ws-6mom-check-wholescale} and \ref{fig:ch-maxent-ws-6mom-check-zoom},
the error is below $20\%$ in most of the domain of interest, except in proximity of the realizability boundary
(Fig.~\ref{fig:ch-maxent-ws-6mom-check-zoom}-Bottom-Right).
Finally, the minimum wave speed can be found from symmetry considerations: $\lambda_\mathrm{min}^\star(q_x^\star) = - \lambda_\mathrm{min}^\star(-q_x^\star)$.

\subsection{Full 14-moment system}

The full 14-moment system is more complex than its axisymmetric counterpart.
Indeed, after non-dimensionalization, the gas state is defined by nine variables:
five components of the dimensionless pressure tensor, $P_{ij}^\star$, three dimensionless heat flux
components, $q_i^\star$, and the parabolic mapping parameter, $\sigma$ (taking the place of $R_{iijj}^\star$).

Some approximations are possible.
First, considering the wave speed along the $x$ direction, one can rotate the reference system around $x$, 
making one of the two heat flux components equal to zero.
In such rotated system, $\tilde{\bm{q}}^\star = (\tilde{q}_x^\star, \tilde{q}_y^\star, 0)$, 
with $\tilde{q}_x^\star \equiv q_x^\star$, and $\tilde{q}_y^\star = (q_y^{\star 2} + q_z^{\star 2})^{1/2}$.
This brings the number of independent variables to eight.
This rotation about $x$ would require that we also rotate the pressure tensor.
However, from numerical inspection, we observed that one can obtain roughly the same wave speeds 
by considering a simpler form of the pressure tensor: isotropic, with an unchanged component along $x$, $\tilde{P}_{xx}^\star = P_{xx}^\star$, and with $\tilde{P}_{yy}^\star = \tilde{P}_{zz}^\star = (P_{yy}^\star + P_{zz}^\star)/2 = (3 - \tilde{P}_{xx}^\star)/2$.

\begin{figure}[h!tb]
  \centering
  \includegraphics[width=1.0\textwidth]{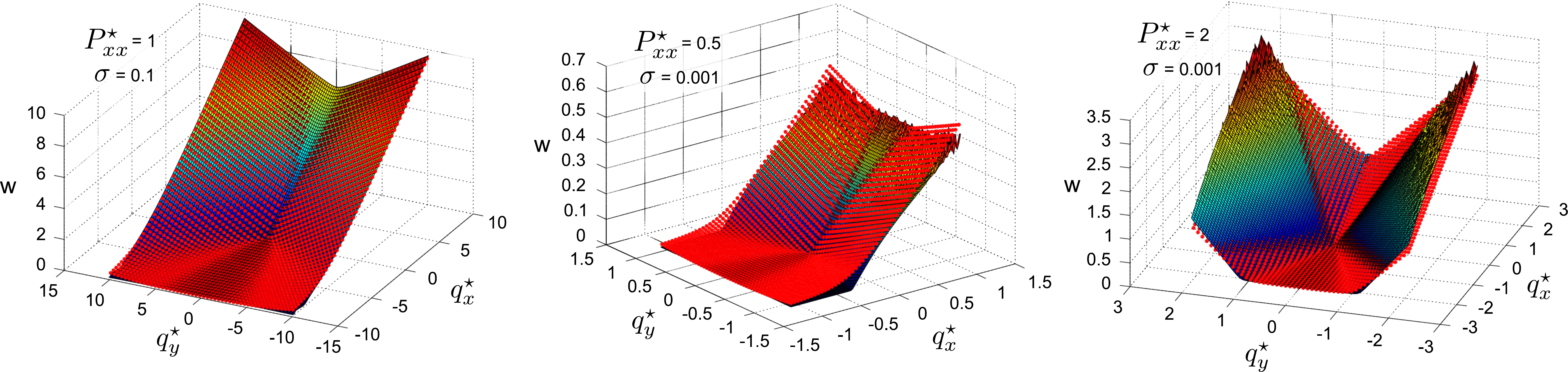}
  \caption{Maximum wave speed of the 14-moment system on the $q_x^\star-q_y^\star$ plane. Surface: numerical eigenvalues.
           Red dots: our approximation, Eq.~\ref{eq:14mom-approx-ws-equations-appendix}. Here, $w = \sigma \lambda_{\mathrm{max}}^\star$.}
  \label{fig:ch-maxent-ws-14mom-qx-qy}
\end{figure}

After these simplifications, the gas state is determined by four quantities, $P_{xx}^\star$, $q_{x}^\star$, $\tilde{q}_y^\star$, and $\sigma$.
This is just one additional quantity with respect to the axisymmetric 14-moment case.
From numerical inspection, we observe that the transverse heat flux, $\tilde{q}_y^\star$, has a strong effect on the
maximum wave speed, as shown in Figure~\ref{fig:ch-maxent-ws-14mom-qx-qy}.
This behavior can be mimicked by embedding the effect of $\tilde{q}_y^\star$ into $q_x^\star$, defining:
\begin{equation}
  \zeta_x = q_x^\star + \alpha \sqrt{\tilde{q}_y^{\star 2}} \, .
\end{equation} 

\noindent In the formulas, we will substitute $\zeta_x$ in place of $q_x^\star$.
Here, $\alpha = \alpha(P_{xx}^\star)$ can be roughly fitted with a parabola,
\begin{equation}
  \alpha(P_{xx}^\star) = 0.6 P_{xx}^{\star 2} - 0.38 P_{xx}^\star + 0.35 \, .
\end{equation}

\noindent The resulting approximation is reasonably good, except for large values of $\sigma$, and in presence of 
a non-zero $\tilde{q}_y$.
In such case, the maximum wave speed is underpredicted, and this might pose some numerical stability issues.
We fix this by introducing a factor $\tilde{q}_y^2/10$ inside the square root, in the hyperbola expression.

Ultimately, the proposed approximation reads
\begin{subequations}\label{eq:14mom-approx-ws-equations-appendix}
\begin{equation}
  \lambda_{\mathrm{max}}^{(14)\star} = \frac{a \sigma + b}{2 \sigma} \left[ \zeta_x + \sqrt{\zeta_x^{2} - 4/5 \zeta_x \sigma C + 4 \sigma^2 Y + \tilde{q}_y^{\star 2}/10} \right] + E  \, ,
\end{equation}
\begin{equation}
  \zeta_x = {q}_x^\star + \sqrt{\tilde{q}_y^{\star 2}} \left( 0.6 \, {P}_{xx}^{\star 2} - 0.38 \, {P}_{xx}^\star + 0.35 \right) \ \  , \ \ \ \tilde{q}_y^\star = (q_y^{\star 2} + q_z^{\star 2})^{1/2} \, ,
\end{equation}
\begin{equation}
  a = 1.4 \, \left({P}_{xx}^\star\right)^{1.1} \ \exp \left[ - {P}_{xx}^{\star 2}\right]  \ , \
  b = 0.9 \, {P}_{xx}^\star \ \exp \left[ - 1/2 \left({P}_{xx}^\star\right)^{1.4}\right] \, ,
\end{equation}
\begin{equation}
   E = \frac{8}{10}\sqrt{(3 - 3\sigma) {P}_{xx}^\star} \ \ \ , \ \ \ \ C = \sqrt{(3 - 3\sigma) {P}_{xx}^\star} \, ,
\end{equation}
\begin{equation}
   B = 5 - 4 \sqrt{\sigma} + \sqrt{10 - 16 \sqrt{\sigma} + 6 \sigma} \ \ \ , \ \ \ 
   Y = B + E^2 - 2 E \sqrt{B} \, .
\end{equation}
\end{subequations}

The \textit{minimum} wave speed is obtained from the maximum wave speed, but reversing the sign of $q_x^\star$, 
leaving the rest unchanged,
\begin{equation}
  \lambda_{\mathrm{min}}^{(14) \star} = -\lambda_{\mathrm{max}}^{(14)\star}(-{q}_x^\star) \, .
\end{equation}

The accuracy of our approximation is shown in Fig.~\ref{fig:ch-maxent-ws-14mom-qx-qy} for some selected cases.
The proposed approximation is much better than the simple estimates based on the
speed of sound alone, that are overly diffisuve in rarefied conditions. 
When necessary, the present wave speeds can be slightly inflated by a multiplication parameter, $k$.
As for the 1D1V 5-moment wave speed approximations, we also observed difficulties in employing these 14-moment 
wave speeds in conjunction with HLL numerical fluxes, when 
rarefied supersonic conditions are simulated, resulting in large numerical oscillations.
Rusanov fluxes instead remain sufficiently smooth and allow one to employ second-order reconstruction techniques.


\section{Euler solution of the crossing jet problem}\label{sec:appendix-euler-jets}

\begin{figure}[h!tb]
  \centering
  \includegraphics[width=1.0\textwidth]{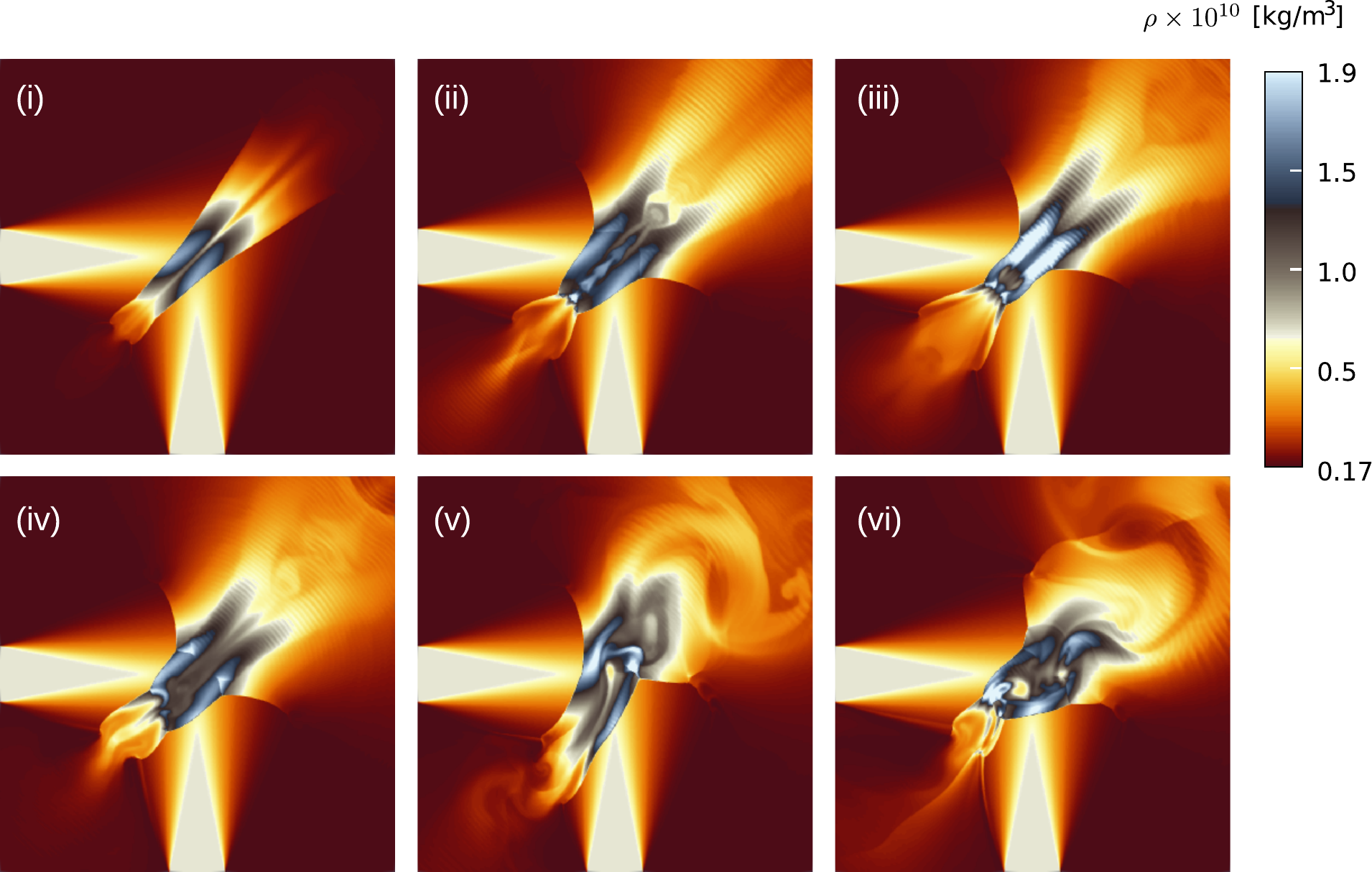}
  \caption{Solution of the Euler equations for the crossing-jet test case of Section~\ref{sec:two-jets-collisionless}, at various time steps.}
  \label{fig:Euler-instability-jets}
\end{figure}

Figure~\ref{fig:Euler-instability-jets} shows the solution of the Euler system for the crossing jets test case.
The computation is performed for a jet density of $\rho_\mathrm{jet}=6.01\times10^{-9}~\si{kg/m^3}$.
Physically, this corresponds to a Knudsen number $\mathrm{Kn} \approx 100$.
However, as expected, the Euler system is unable to reproduce jets re-separation, regardless of the low density. 
Moreover, the system appears to develop a number of shock waves.
This evolves first into a symmetric instability (panels i–iii), and eventually switches to an asymmetric mode (panels iv–vi).

\end{document}